\documentclass[pdflatex,sn-basic,twocolumn]{sn-jnl}

\usepackage{graphicx} 
\usepackage{multirow}%
\usepackage{amsmath,amssymb,amsfonts}%
\usepackage{amsthm}%
\usepackage{mathrsfs}%
\usepackage[title]{appendix}%
\usepackage{xcolor}%
\usepackage{textcomp}%
\usepackage{manyfoot}%
\usepackage{booktabs}%
\usepackage{algorithm}%
\usepackage{algorithmicx}%
\usepackage{algpseudocode}%
\usepackage{listings}%
\usepackage{caption}
\usepackage{booktabs}
\usepackage{float}
\usepackage{placeins}
\usepackage{rotating}

\theoremstyle{thmstyleone}%
\theoremstyle{thmstyletwo}%

\theoremstyle{thmstylethree}%
\begin{document}

\title[Article Title]{The long-term optical flux variations of Compact Symmetric Objects }

\title{The long-term optical flux variations of Compact Symmetric Objects 
}


\author*[1]{\fnm{Subhashree} \sur{Swain}}\email{s.subhashree00@gmail.com}

\author[1]{\fnm{Vaidehi S.} \sur{ Paliya}}\email{Vaidehi.s.paliya@gmail.com}

\author[1]{\fnm{D. J.} \sur{Saikia}}\email{dhrubasaikia.tifr.ccsu@gmail.com}
\author[2]{\fnm{C. S.} \sur{Stalin}}\email{stalin@iiap.res.in}
\author[3]{\fnm{Arya} \sur{Venugopal}}\email{aryakaruppal02@gmail.com}
\author[3]{\fnm{A. K.} \sur{Bhavya}}\email{bhavyaak42@gmail.com}
\author[3]{\fnm{C. D.} \sur{Ravikumar}}\email{cdr@uoc.ac.in}


\affil*[1]{\orgaddress{Inter-University Centre for Astronomy and Astrophysics (IUCAA), SPPU Campus, Ganeshkhind, Pune 411007, India }}

\affil[2]{\orgaddress{Indian Institute of Astrophysics, Block II, Koramangala, Bangalore 560034, India}}

\affil[3]{\orgaddress{Department of Physics, University of Calicut, Malappuram 673635, India}}



\abstract{Compact Symmetric Objects (CSOs) are a distinct category of jetted active galactic nuclei (AGN) whose optical variability characteristics have not been well investigated. We present here the results of our investigation on the optical flux and colour variability properties of a bona fide sample of 38 CSOs. We used the $g$-, $r$- and $i$-bands data from the Zwicky Transient Facility survey that spans a duration of about 5 years. 
We also considered a comparison sub-sample of blazars that includes 5 flat spectrum radio quasars and 12 BL Lac objects with redshifts and $g$-band magnitudes similar to the limited sub-sample of 9 CSOs.
These two sub-samples of AGN, chosen for this comparative study of their long-term optical variability, represent different orientations of their relativistic jets with respect to the observer. 
We found that both CSOs and blazars exhibit optical flux variations, although variability of CSOs is lower than that of blazars.
The observed variability in both CSOs and blazars is attributed to the relativistic jets and the increased optical variations in blazars relative to CSOs are likely due to beaming effects. CSOs and blazars exhibit similar colour variations, with both of them showing a bluer when brighter trend. Such a colour variability pattern is expected due to processes associated with their relativistic jets.}



\maketitle

\section{Introduction}
Compact symmetric objects (CSOs) are a distinct category of jetted active
galactic nuclei (AGN) identified by their overall projected linear radio sizes smaller than 1 kpc \citep[e.g.,][]{Kiehlmann2023b,Kiehlmann2023}.
They are compact with their oppositely-directed jets forming lobes/hotspots with an overall projected linear size $<$ 1 kpc. These sources tend to have an inverted radio spectrum and constitute a subclass of the peaked-spectrum radio sources \citep{Wilkinson1994, readhead1978,readhead1980,readhead1996,ODea2021}. They  
are thought to be young radio sources with kinematic ages between $\sim$100 and 3000 yr estimated from the velocity of the outer hotspots with the radiative age estimates being consistent with their young nature \citep[cf.][]{Taylor2000, an2012,ODea2021}. With steep high-frequency spectra and relatively weak cores and bright hotspots, the CSOs are expected to lie at larger angles to the line of sight compared with the core-jet sources, so that the effects of relativistic beaming are small \citep[e.g.][]{Krezinger2020}. However, a few edge-dimmed CSOs have also been identified \citep[e.g.,][]{Kiehlmann2023b}.

The small sizes of compact steep-spectrum sources including CSOs and peaked-spectrum radio sources have been traditionally attributed to (i) their young ages so that they are precursors of the larger radio galaxies and quasars; (ii) a dense galactic environment that confines their jets to small sizes; and (iii) transient or recurrent jet activity. 
There have been recent suggestions that CSOs represent transient jet activity and they do not evolve into larger sources \citep{Readhead2023, Kiehlmann2023b, Kiehlmann2023}.
Some CSOs are also found to exhibit asymmetric radio structure, possibly due to differences in the density of the surrounding medium \citep[e.g.][]{Saikia1995MNRAS.276.1215S,1996ApJ...463...95T}. 
A high degree of polarization asymmetry of the outer lobes compared to the conventional Fanaroff-Riley type I and II radio sources is also consistent with a dense asymmetric environment in the central regions of CSO host galaxies \citep[cf.][]{SaikiaGupta2003AnA...405..499S}. Also, X-ray observations have provided evidence for the existence of a dense medium in a few CSOs with the absorbing neutral hydrogen column density exceeding 10$^{23}$ cm$^{-2}$ \citep{2019ApJ...884..166S,2019ApJ...871...71S,2023ApJ...948...81S}. Neutral atomic hydrogen $\lambda$21-cm observations show an inverse correlation of column density with projected linear size, with the CSOs exhibiting the highest values \citep[][]{2003A&A...404..871P,2006MNRAS.373..972G}.
Some of the CSOs have also been detected in the GeV band by the Fermi-Large Area Telescope. The CSOs that are known to be emitters of GeV photons are PKS 1718$-$649 \citep{2016ApJ...821L..31M}, NGC 3894 \citep{2020A&A...635A.185P}, TXS 0128+554 \citep{2020ApJ...899..141L}, and DA 362 \citep[][]{2024arXiv241212857S}.

One of the defining characteristics of an AGN is the detection of flux variations
across the electromagnetic spectrum on diverse time scales \citep[minutes-to-years;][]{1995ARA&A..33..163W,1997ARA&A..35..445U}. The analysis of the flux variability characteristics provides important clues about the physical processes responsible for the observed emission and their possible interaction with the surrounding environment. Among AGN population, blazars, comprising of flat spectrum radio quasars (FSRQs) and BL Lac objects (BL Lacs), tend to exhibit the strongest flux variability at all wavelengths \citep[e.g.,][]{1996ApJ...461..698H,2006MNRAS.366.1337S,2015ApJ...803...15P,2015ApJ...811..143P,2017ApJ...835..275R,2017MNRAS.466.3309R,2020A&A...634A..80R,2023MNRAS.526.4502R}. They show
extreme luminosities up to about 10$^{47-48}$ erg s$^{-1}$  and their radiation output is dominated by the Doppler-boosted non-thermal jet emission \citep[e.g.,][]{2014Natur.515..376G}. 
The high variability seen in blazars is
generally explained by the shock-in-jet model \citep{1985ApJ...298..114M} 
and the large variability amplitude identified at short timescales in the optical band is attributed to relativistic beaming \citep[cf.][]{2003ApJ...586L..25G}.
In addition to temporal flux variations, blazars also exhibit spectral variability and
detailed investigations of optical colour variations have been carried out to reveal the origin of such behaviour \citep[e.g.][]{2012ApJ...756...13B,2019MNRAS.484.5633G,2022ApJS..259...49Z,
2023MNRAS.519.5263Z}. Two patterns of colour variations are generally observed, namely (i) a bluer when brighter (BWB) behaviour, in which the source becomes bluer (spectrum becomes harder) when it brightens 
\citep{2009MNRAS.399.1357S,2019AJ....157...95G}  and 
(ii) a redder when brighter (RWB) behaviour in which the source becomes
redder (spectrum becomes softer) when it brightens \citep{2017Natur.552..374R}.

Different categories of AGN have been studied for flux and spectral variations in different wavelengths; however, we have little or no information on the optical variability characteristics of CSOs. The main objective of this work is to characterise the optical flux
variability properties of a bona fide sample of CSOs. 
Also, observations in the radio band indicate that for compact steep spectrum sources (which include CSOs), the ones associated with quasars are statistically inclined at smaller angles to the line of sight than those associated with galaxies \citep[e.g.][]{Saikia1995MNRAS.276.1215S,ODea2021}. Therefore, if jets and beaming are primarily responsible for the observed variability, one might expect higher levels of flux variations in CSOs associated with quasars compared to the ones associated with galaxies.
In this work, we aim to (a) characterise the optical flux variability of a sample of CSOs, (b) investigate how the optical flux variations of CSOs compare with those of blazars, and (c) look for similarities and/or differences in the spectral variability behaviour of CSOs and blazars. The structure of the paper is as follows. The selection
of the samples of CSOs and blazars, as well as the data used in this work, are described in Section 2, while the analysis and results
are presented in Section 3. The results are discussed in Section 4, 
which is followed by the conclusions in Section 5.

\begin{table*}
\centering
\caption[]{Details of the sources studied in this work. Columns show (1) the J2000 name, (2) a common name, (3,4) J2000 right ascension and declination, (5) redshift, $z$, (6) SDSS $g$-band magnitude, (7) optical classification: `Q' stands for quasar, `G' stands for galaxy. $^a$ - photometric redshift from \cite{Bilicki_2014}.}
\label{table-1}
\fontsize{8pt}{10pt}\selectfont
\begin{tabular}{clccccc} 
\hline
Name &  Other name & RA & Dec & z & $g$-band & Type \\
     &             &    &     &   &  (mag) &\\
     (1) & (2) & (3) & (4) & (5) & (6)  & (7) \\
\hline
\hline
\multicolumn{6}{c}{CSOs}\\
J0119+3210 & B2~0116+31    & 01:19:35.00 & +32:10:50.06 &  0.060 & 17.49 & G\\
J0131+5545 & TXS 0128+554 & 01:31:13.82 & +55:45:12.98 &  0.036 &18.44  &G\\
J0402+8241 & JVAS J0402+8241 & 04:02:12.68& +82:41:35.13 &  0.066$^a$      &17.69 &  G\\
J0405+3803 & B3 0402+379 & 04:05:49.26& +38:03:32.24 &  0.056 & 21.59&G\\
J0713+4349 & B3 0710+439 & 07:13:38.16& +43:49:17.21 &  0.518 &21.93 & G\\
J0741+2706 & B2 0738+27 & 07:41:25.73& +27:06:45.42 &  0.772 &19.92 &Q \\
J0832+1832 & PKS 0829+18 & 08:32:16.04& +18:32:12.12 &  0.154 &18.80 & G\\
J0855+5751 & JVAS J0855+5751 & 08:55:21.36& +57:51:44.09 &  0.026 &21.91&G\\
J0906+4124 & GB6 J0906+4124 & 09:06:52.80& +41:24:30.00 &  0.027 &16.51 &G\\
J0909+1928 & MRK 1226 & 09:09:37.44& +19:28:08.30 &  0.028 & 16.40&G\\
J1025+1022 & NVSS J102544+102231 & 10:25:44.20& +10:22:30.00 &  0.046 &17.21&G\\
J1035+5628 & JVAS J1035+5628 & 10:35:07.04& +56:28:46.79 &  0.460 &22.13& G\\
J1111+1955 & PKS 1108+201 & 11:11:20.07& +19:55:36.01 &  0.299 &20.36& G\\
J1120+1420 & PKS 1117+146 & 11:20:27.81& +14:20:54.97 &  0.362 &21.75&  G\\
J1148+5924 & NGC 3894 & 11:48:50.36& +59:24:56.36 &  0.011 &15.29 & G\\
J1158+2450 & PKS 1155+251 & 11:58:25.79& +24:50:18.00 &  0.203 &19.48 & G\\
J1159+5820 & VERA J1159+5820 & 11:59:48.77& +58:20:20.31 &  1.280 &21.33 &G\\
J1205+2031 & NGC 4093 & 12:05:51.50& +20:31:19.00 &  0.024 &16.44&G\\
J1220+2916 & NGC 4278 & 12:20:06.82& +29:16:50.72 &  0.002 &14.05 & G\\
J1227+3635 & B2~1225+36 & 12:27:58.72& +36:35:11.82 &  1.975 &21.96&  Q\\
J1234+4753  & JVAS J1234+4753 & 12:34:13.33& +47:53:51.24 &  0.373 &17.21&  Q\\
J1244+4048 & B3 1242+410 & 12:44:49.19& +40:48:06.15 &  0.814 &20.25& Q\\
J1254+1856 & CRATES J1254+1856 & 12:54:33.27& +18:56:01.93 &  0.115 &18.63&  G\\ 
J1326+3154 & DA 344 & 13:26:16.51 &+31:54:09.52 &  0.368 &21.04& G\\
J1407+2827 & OQ 208 & 14:07:00.40& +28:27:14.69 &  0.077 & 16.36& G\\
J1508+3423 & VV 059a & 15:08:05.70 &+34:23:23.00 &  0.046 &18.03 &G\\
J1511+0518 & JVAS J1511+0518 & 15:11:41.27& +05:18:09.26 &  0.084 &17.67 & G\\
J1559+5924 & JVAS J1559+5924 & 15:59:01.70& +59:24:21.84 &  0.060 &16.77& Q\\
J1602+5243 & 4C +52.37 & 16:02:46.38& +52:43:58.40 &  0.106 &17.76&  G \\
J1609+2641 & CTD 93 & 16:09:13.32& +26:41:29.04 &  0.473 &22.27 &  G\\
J1816+3457 & B2 1814+34 & 18:16:23.90& +34:57:45.75 &  0.245 &20.52 &G \\
J1915+6548 & JVAS J1915+6548 & 19:15:23.82& +65:48:46.39 &  0.486 &19.53&  Q\\
J1928+6815 & JVAS J1928+6814 & 19:28:20.55& +68:14:59.27 &    -    &21.09&-\\
J1944+5448 & S4 1943+54 & 19:44:31.51& +54:48:07.06 &  0.263 &20.88 &  G\\
J1945+7055 & S5 1946+70 & 19:45:53.52& +70:55:48.73 &  0.101 &19.31 & G \\
J2022+6136 & S4 2021+61 & 20:22:06.68 &+61:36:58.80 &  0.227 &19.82 & G\\
J2327+0846 & NGC 7674 & 23:27:56.70& +08:46:44.30 &  0.029 &15.71 &G \\
J2355+4950 & TXS 2352+495 & 23:55:09.46& +49:50:08.34 &  0.238 &20.71 & G\\
\hline
\multicolumn{6}{c}{FSRQs}\\
J0117+1418 & BZQ~J0117+1418&  01:17:25.20	& +14:18:12.38 & 0.839 &20.83 & \\
J0914+0245 &  PKS~0912+029& 09:14:37.91 & +02:45:59.18 & 0.427 & 19.15 &\\
J1140+4622 & Q 1138+4638&  11:40:47.90  & +46:22:04.90  & 0.115 & 16.75 &\\
J1143+1843 & RGB J1143+187&  11:43:06.03	& +18:43:42.81 & 0.374 &16.83 & \\
J1716+2152 &GC 1714+21&  17:16:11.18	& +21:52:13.69 & 0.358 &19.92 &\\			
\hline 
\multicolumn{6}{c}{BL Lacs}\\
J0006$-$0623&PKS~0003-066	&  00:06:13.89	& $-$06:23:35.19	& 0.347 &18.98 & \\
J0127$-$0821& FBQS J0127$-$0821 &  01:27:16.31	 & 
$-$08:21:28.90	& 0.362  & 19.20&\\
J0809+5218& 1ES~0806+524	&  08:09:49.19	&+52:18:58.39	& 0.138 &15.68  & \\
J0847+1133&RXS~J08472+1133	&  08:47:12.93	&+11:33:50.29	& 0.199 &18.46 & \\
J0909+2311& SDSS~J09089+2311	&  09:09:00.61	&+23:11:12.91	& 0.223 &18.06 &\\
J1058+5628&RX~J10586+5628	&  10:58:37.73	 & +56:28:11.20	& 0.143 &15.48 & \\
J1104+3812&MKN~421	&  11:04:27.30	& +38:12:31.78	& 0.030  &13.24 &\\
J1203+6031 &SBS 1200+608	&  12:03:03.50	& +60:31:19.09	& 0.065 &16.45  &\\
J1221+2813 &ON 231 &  12:21:31.68	 & +28:13:58.51 & 0.102 & 15.13 &\\
J1423+1412&BZB J1423$+$1412	&  14:23:30.66	&+14:12:47.80	& 0.769 &19.58&\\
J1735+5650&BZB~J1735+5650	&  17:35:10.81	&+56:50:22.70	& 0.388 &21.74 &\\
J2357$-$0152&PKS 2354-021	&  23:57:25.14 & $-$01:52:15.49	& 0.812 &18.78 &\\
\hline
\end{tabular}
\end{table*}

\begin{table*}
\centering
\caption[]{Variability amplitudes of the sources studied in this work. Columns 2, 3, and 4 refer to the number of photometric points in $g$, $r$ and $i$ bands, respectively, and 5, 6, and 7 correspond to the fractional variability amplitude (F$_{var}$) in $g$, $r$ and $i$ bands. The columns 8, 9, and 10 report the probability of the source being variable, estimated based on the $\chi^2$ test in all three bands. `V' stands for Variable, `PV' stands for probable variable and `NV' stands for non-variable.}
\label{table2}
\tiny
\begin{tabular}{crrrllllll} 
\hline
Name & \multicolumn{3}{c}{No. of data points} & F$_{var,g}$ &F$_{var,r}$ &F$_{var,i}$ & P$_{v,g}$ &P$_{v,r}$&P$_{v,i}$ \\
     &  $g$  & $r$ & $i$ &  &  &\\
    (1) & (2) & (3) & (4) & (5) & (6) & (7) & (8) & (9) & (10) \\
\hline
\multicolumn{10}{c}{CSOs}\\
J0119+3210  & 742  & 966&100&0.107$\pm$0.001  & 0.076$\pm$0.001 &0.081$\pm$0.001& >99\% (V)& >99\% (V)& >99\% (V)\\
J0131+5545 & 405  &716 &-  & 0.092$\pm$0.002 &0.054$\pm$0.005  &-&>99\% (V)&>99\% (V)&-\\
J0402+8241 & 44&47&-&0.110$\pm$0.004&0.067$\pm$0.002&-&>99\% (V)&>99\% (V)&-\\
J0405+3803 & 14  & 532& 12&0.118$\pm$0.064&0.138$\pm$0.003&0.099$\pm$0.011&<95\% (NV)&>99\% (V)&>99\% (V)\\
J0713+4349 & 9  &553& 36&0.187$\pm$0.093&0.095$\pm$0.011&0.059$\pm$0.036 &>99\% (V)&>99\% (V)&<95\% (NV)\\
J0741+2706 &  34&28&-&0.240$\pm$0.013&0.239$\pm$0.019&-&>99\% (V)&>99\% (V) &-\\
J0832+1832 & 264 &716&38&0.081$\pm$0.003&0.044$\pm$0.001&0.043$\pm$0.003& >99\% (V)& >99\% (V)& >99\% (V)\\
J0855+5751 &  15  &311&30&0.093$\pm$0.133&0.027$\pm$0.063&0.228$\pm$0.031&<95\% (NV)&<95\% (NV)&>99\% (V)\\
J0906+4124 & 449  &555&45&0.093$\pm$0.001&0.063$\pm$0.001&0.061$\pm$0.002&>99\% (V)& >99\% (V)& >99\% (V)\\
J0909+1928 & 229 &684&36&0.066$\pm$0.001&0.046$\pm$0.001&0.051$\pm$0.002&>99\% (V)& >99\% (V)& >99\% (V)\\
J1025+1022 & 204&339&32&0.133$\pm$0.001&0.088$\pm$0.001&0.067$\pm$0.002&>99\% (V)& >99\% (V)& >99\% (V)\\
J1035+5628 & 11&332&42&-&0.101$\pm$0.018&- &<95\% (NV)&>99\% (V)& <95\% (NV)\\
J1111+1955 & 130&312&41&0.037$\pm$0.039&0.069$\pm$0.005&0.050$\pm$0.010 &<95\% (NV)&>99\% (V)& >99\% (V)\\
J1120+1420 &  14&179&100&0.244$\pm$0.065&-&0.041$\pm$0.030 &>99\% (V)&<95\% (NV)&<95\% (NV)\\
J1148+5924  & 663&713&157&0.118$\pm$0.001&0.079$\pm$0.001&0.079$\pm$0.001 &>99\% (V)& >99\% (V)& >99\% (V)\\
J1158+2450  & 445&651&167&0.075$\pm$0.006&0.062$\pm$0.002&0.046$\pm$0.002 & >99\% (V)& >99\% (V)& >99\% (V)\\
J1159+5820 & 190&372 &86&0.034$\pm$0.082&-&0.067$\pm$0.037&<95\% (NV)&<95\% (NV)&<95\% (NV)\\
J1205+2031 & 322&404&124&0.067$\pm$0.001&0.062$\pm$0.001&0.069$\pm$0.001&>99\% (V)& >99\% (V)& >99\% (V)\\
J1220+2916 & 510&638&172&0.089$\pm$0.001&0.065$\pm$0.001&0.085$\pm$0.001 &>99\% (V)& >99\% (V)& >99\% (V) \\
J1227+3635 &  35&99&36&0.148$\pm$0.056&0.089$\pm$0.044&0.143$\pm$0.046 & <95\% (NV)&<95\% (NV)&>95\% (PV)\\
J1234+4753 & 552&646&189&0.029$\pm$0.001&0.028$\pm$0.001&0.022$\pm$0.001 & >99\% (V)& >99\% (V)& >99\% (V)\\
J1244+4048 & 477&595&155&0.197$\pm$0.008&0.230$\pm$0.006&0.188$\pm$0.010 &>99\% (V)& >99\% (V)& >99\% (V)\\ 
J1254+1856 &  28&34&-&0.199$\pm$0.011&0.061$\pm$0.004&-&>99\% (V)& >99\% (V)& -\\ 
J1326+3154 & 188&485&183&0.080$\pm$0.025&-&0.058$\pm$0.007& <95\% (NV)&<95\% (NV)&>99\% (V)\\
J1407+2827 & 342&437&104&0.087$\pm$0.001&0.084$\pm$0.001&0.064$\pm$0.001 & >99\% (V)& >99\% (V)& >99\% (V)\\
J1508+3423 &  16&765&145&0.082$\pm$0.007&0.165$\pm$0.001&0.123$\pm$0.001&>99\% (V)& >99\% (V)& >99\% (V)\\
J1511+0518 &267&452&617&0.052$\pm$0.002&0.045$\pm$0.001&0.056$\pm$0.002 &>99\% (V)& >99\% (V)& >99\% (V)\\
J1559+5924 & 1103&1102&353&0.067$\pm$0.001&0.059$\pm$0.001&0.044$\pm$0.001 & >99\% (V)& >99\% (V)& >99\% (V)\\
J1602+5243 & 955&1047&144&0.067$\pm$0.001&0.061$\pm$0.001&0.083$\pm$0.001& >99\% (V)& >99\% (V)& >99\% (V)\\
J1609+2641 &  15&310&109&-&0.100$\pm$0.013&0.020$\pm$0.048&<95\% (NV)& >99\% (V)& <95\% (NV)\\
J1816+3457 & 535&1102&90&0.089$\pm$0.008&0.048$\pm$0.002&0.022$\pm$0.009 & >99\% (V) & >99\% (V)& >99\% (V) \\
J1915+6548 & 897&927&107&0.054$\pm$0.003&0.031$\pm$0.003&0.039$\pm$0.005 & >99\% (V)& >99\% (V)& >99\% (V) \\
J1928+6815 & 328& 843&102&0.131$\pm$0.014&0.054$\pm$0.005&0.006$\pm$0.078& >99\% (V)& >99\% (V)& <95\% (NV)\\
J1944+5448 &  13&63&-&0.096$\pm$0.071&0.110$\pm$0.009&-& <95\% (NV)& >99\% (V)& -\\
J1945+7055 & 471&527&-&0.096$\pm$0.003&0.064$\pm$0.001&- & >99\% (V)& >99\% (V)& -  \\
J2022+6136 &750&1022&113&0.063$\pm$0.004&0.049$\pm$0.001&0.049$\pm$0.003&  >99\% (V)& >99\% (V)& >99\% (V)\\
J2327+0846 & 55&77&29&0.036$\pm$0.002&0.046$\pm$0.001&0.056$\pm$0.003 & >99\% (V)& >99\% (V)& >99\% (V) \\
J2355+4950 & 429&954& 87&0.105$\pm$0.008&0.052$\pm$0.002&0.056$\pm$0.005 & >99\% (V)& >99\% (V)& >99\% (V)\\
\hline
\multicolumn{10}{c}{FSRQs}\\
J0117+1418  &212 & 407   & 58 &0.462$\pm$0.008&0.413$\pm$0.005&0.395$\pm$0.013 & >99\% (V)& >99\% (V)& >99\% (V) \\
J0914+0245 &  137   &  558  &  31&0.342$\pm$0.009&0.257$\pm$0.004&0.119$\pm$0.015& >99\% (V)& >99\% (V)& >99\% (V)  \\
J1140+4622 &  531 & 602 & 173 &0.055$\pm$0.001&0.042$\pm$0.001&0.029$\pm$0.001& >99\% (V)& >99\% (V)& >99\% (V)\\
J1143+1843 &    292 & 412   & 129 &0.072$\pm$0.001&0.068$\pm$0.001&0.025$\pm$0.002 & >99\% (V)& >99\% (V)& >99\% (V)\\
J1716+2152 &    280 &  605  & 155 &0.253$\pm$0.006&0.16$1\pm$0.002&0.111$\pm$0.004 & >99\% (V)& >99\% (V)& >99\% (V) \\			
    \hline 
    \multicolumn{10}{c}{BL Lacs}\\
J0006$-$0623 &235 & 285 & 40 &0.230$\pm$0.005&0.196$\pm$0.003&0.263$\pm$0.008 & >99\% (V)& >99\% (V)& >99\% (V)\\
J0127$-$0821& 232   & 270   &  - &0.305$\pm$0.004&0.291$\pm$0.003&- & >99\% (V)& >99\% (V)& -\\
J0809+5218 & 537  & 679   & 47 &0.120$\pm$0.001&0.117$\pm$0.001&0.086$\pm$0.002 & >99\% (V)& >99\% (V)& >99\% (V) \\
J0847+1133 & 270 &559    & 38 &0.215$\pm$0.002&0.157$\pm$0.001&0.125$\pm$0.003 & >99\% (V)& >99\% (V)& >99\% (V)\\
J0909+2311 &  273    &522  & 39 &0.210$\pm$0.001&0.217$\pm$0.001&0.173$\pm$0.003 & >99\% (V)& >99\% (V)& >99\% (V) \\
J1058+5628 &254  &  709  &  85 &0.153$\pm$0.001&0.154$\pm$0.001&0.148$\pm$0.002 & >99\% (V)& >99\% (V)& >99\% (V)\\
J1104+3812 & 516 & 555   & 47 &0.251$\pm$0.001&0.222$\pm$0.001&0.211$\pm$0.002  & >99\% (V)& >99\% (V)& >99\% (V)\\
J1203+6031 & 584 & 722   & 158  &0.141$\pm$0.001&0.105$\pm$0.001&0.087$\pm$0.001 & >99\% (V)& >99\% (V)& >99\% (V)\\
J1221+2813 & 290 & 409 & 13 &0.535$\pm$0.001&0.487$\pm$0.001&0.316$\pm$0.001  & >99\% (V)& >99\% (V)& >99\% (V) \\
J1423+1412 & 256    &  390  & 85 &0.134$\pm$0.006&0.134$\pm$0.004&0.083$\pm$0.010& >99\% (V)& >99\% (V)& >99\% (V)  \\
J1735+5650 & 64    & 980   & 162 &0.249$\pm$0.034&0.045$\pm$0.010&0.049$\pm$0.018 & >99\% (V)& >95\% (PV)& <95\% (NV) \\
J2357$-$0152 &  197   &  259  &  75&0.517$\pm$0.005&0.486$\pm$0.003&0.346$\pm$0.005 & >99\% (V)& >99\% (V)& >99\% (V) \\
\hline
\end{tabular}
\end{table*}

\section{Sample and data}
\subsection{CSOs}
We considered the catalog of 79 bona fide CSOs recently published by \citet{Kiehlmann2023b}. The catalog was prepared by searching the literature and by analyzing their multi-band radio observations of the potential CSOs. The following four criteria were adopted to select objects: (i) projected source length smaller than 1 kpc, (ii) detection of bi-polar radio emission, (iii) no significant radio flux variability, and (iv) no superluminal motion in excess of $v_{\rm app}=2.5c$. 
For the 79 sources, we searched for the availability of multi-band optical data in the Zwicky Transient Facility (ZTF; \citealt{Graham2019}) 
database\footnote{https://www.ztf.caltech.edu/ztf-public-releases.html}.
The ZTF survey scans the northern sky in $g$ (4829.5 \AA), 
$r$ (6463.75 \AA), and $i$ (7903.55 \AA) bands every 2$-$3 days with an average exposure length of 30 seconds. It uses a wide-field camera with a field of view of 47 deg$^2$ mounted on the 48-inch Samuel Oschin Schmidt telescope \citep{Bellm2019}. The ZTF treats the light curves observed in a particular field, filter, and CCD quadrant independently and assigns a different observation ID to the source observed in different combinations of the three. These CCD quadrants are calibrated independently
as this could give spurious variability after combining light curves from
different fields and CCD quadrants \citep{vanRoestel_2021}.
Thus, to avoid spurious variability from the different observation IDs, we 
only took the light curve corresponding to the observation ID with the maximum 
number of data points. In the ZTF database, there are 38 sources, 6 of those having observations in $g$ and $r$ bands only and 32 having observations in $g$, $r$, and $i$ bands. Of the 38 CSOs, 36 sources have redshift information from \cite{Kiehlmann2023b}, one source has redshift information from \cite{Bilicki_2014} and one source has no redshift measurement. We adopted their SDSS $g$- and $r$-band magnitudes and calculated the apparent $B$-band magnitude using the following relation given by \cite{jordi2006}.
\begin{equation}
\small
B = g + (0.313\pm0.003) \times (g-r) + (0.219 \pm 0.002)
\end{equation}
We then calculated the k-corrected absolute magnitude following the method adopted in the SDSS quasar catalog \citep[][]{2020ApJS..250....8L}. Sources less luminous than M$_B$ of $-$23 were classified as galaxies, and those more luminous than $-$23 were classified as quasars \citep{smdt1983}. Based on this criterion, out of the 38 CSOs, 6 and 31 are identified as quasars and galaxies, respectively. One object could not be classified due to lack of the redshift information.

\subsection{Blazars}
To make a comparative analysis of the flux variability characteristics of CSOs and blazars, we selected a sample of blazars matched in redshift and optical brightness of the CSOs. The blazars were chosen from the ROMA-BZCAT\footnote{https://www.ssdc.asi.it/bzcat5/} catalog that contains 3561 sources \citep{2015Ap&SS.357...75M}. For the 38 CSOs with available ZTF photometry, we searched for FSRQs and BL Lacs that match with the redshift of the CSOs within $\pm$0.05 and with the $r$-band magnitude within $\pm$ 0.5 mag. For some CSOs, we did not have a matched FSRQ and for some CSOs, we have more than one matched BL Lac. This resulted in a final sample of 9 CSOs, 5 FSRQs and 12 BL Lacs, which forms our limited sub-sample of CSOs and blazars for a comparative study in all three bands. The reference $g$-band magnitude of the selected sample spans from 13.24 to 21.75 mag, while the corresponding redshift ranges from 0.024 to 0.839.

\begin{figure}[h!]
\centering
     \includegraphics[width=\columnwidth]{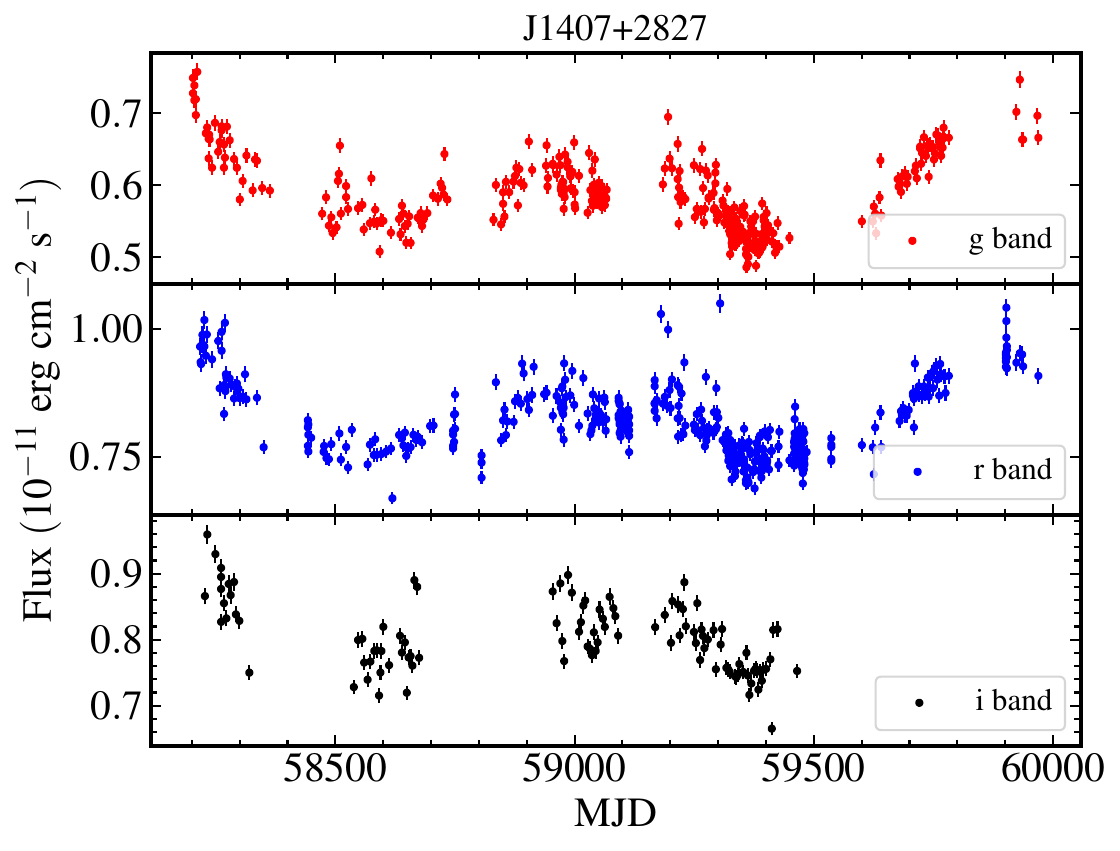}
     \includegraphics[width=\columnwidth]{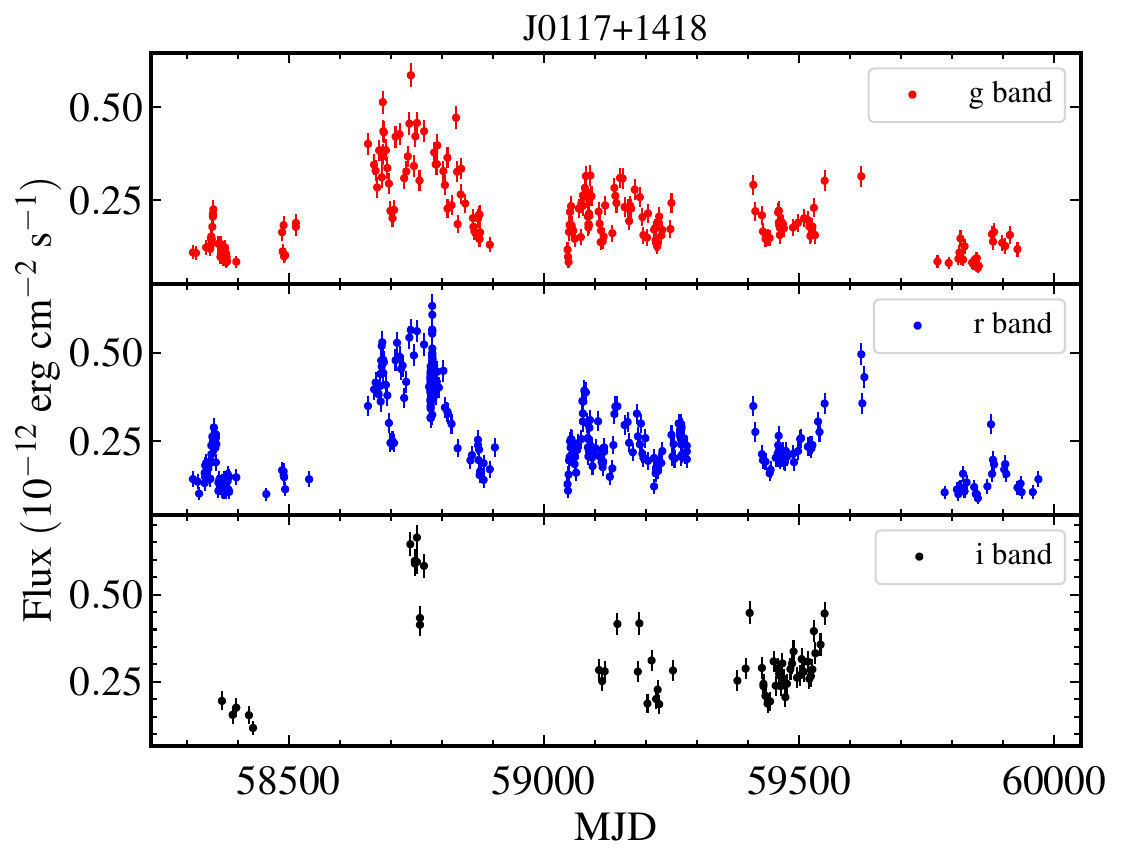}
     \includegraphics[width=\columnwidth]{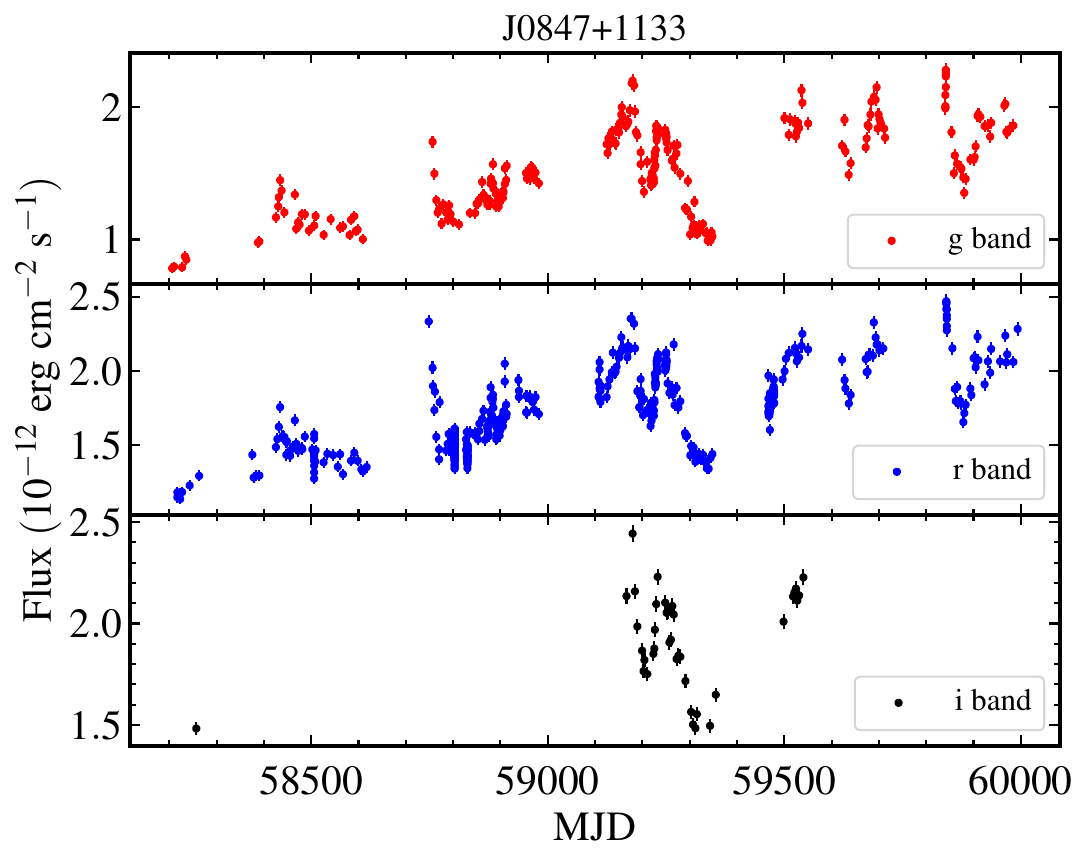}
    \caption{The light curves in $g$, $r$ and $i$ bands for the CSO J1407+2827 (top panel), the  FSRQ J0117+1418 (middle panel) and the BL Lac J0847+1133 (bottom panel).}
\label{figure-1}
\end{figure}

To have clean light curves of those 38 CSOs, 5 FSRQs and 12 BL Lacs, we used only those photometric points having {\sc catflags} = 0, i.e., which are obtained in good observing conditions. Also, to remove deviant outliers from the light curves, we used a 3$\sigma$ clipping method. The basic information of the 38 CSOs along with the number of epochs for which good observations are available, are
given in Table \ref{table-1}. Similarly, the details of the FSRQs and BL Lacs that are used for the comparative analysis are given in Table \ref{table-1}. The 9 CSOs that are used for comparative analysis are J0832+1832, J1111+1955, J1120+1420, J1158+2450, J1205+2031, J1234+4753, J1244+4048, J1326+3154, and J1559+5924. Finally, we corrected the apparent magnitudes for the Galactic extinction following \citet[][]{2011ApJ...737..103S} and converted them to energy flux units \citep[][]{1983ApJ...266..713O}. In Figure \ref{figure-1}, we show the multi-band light curves of a CSO, a FSRQ, and a BL Lac object.
 \section{Analysis and Results}
\subsection{Temporal Variability}
\subsubsection{Chi-square test}
We investigated whether a source has significant variations using a $\chi^2$ test, given by

\begin{equation}
\chi^2 = \sum_{i=1}^{N} \frac{(f_i - \bar{f_i})^2}{f_{err,i}^2}
\end{equation}

where $f_i$ represents the flux measurement in the $i^{th}$ time bin, $f_{err,i}$ is the uncertainty associated with each measurement, and $\bar{f_i}$ is the mean of $f_i$ for the source. Here, $N$ denotes the number of points in each light curve.
If the computed $\chi^2$ exceeded the critical $\chi^2$ value ($\chi^2_{\text{critical}}$) at $\gtrsim$99\% confidence level (c.l.), the source is classified as variable (V). If the estimated $\chi^2$ lied between $\chi^2_{\text{critical,99\%~c.l.}}$ and $\chi^2_{\text{critical,95\%~c.l.}}$, the source is classified as a probable variable (PV). The sources for which the $\chi^2$ test does not exceed $\chi^2_{\text{critical}}$ at 95\% c.l., were classified as non-variable (NV).

On the basis of the $\chi^2$ test, variability was detected in 21 out of 38 CSOs in all $g$, $r$, and $i$ bands. Considering individual bands, 29 (9) CSOs were found variable (non-variable) in the $g$ band, and 33 (5) CSOs were identified as variable (non-variable) in the $r$ band. Considering the $i$-band data, 25 CSOs were variable, 6 were non-variable, and one has shown hints of flux variations, i.e., probable variable. 
Our CSO sample contains two $\gamma$-ray emitting sources. One is J0131$+$5545 (TXS 0128$+$554), which showed variability in the $g$ and $r$ bands, 
while the other J1148$+$5924 (NGC 3894), exhibited variability in the $g$, $r$ and $i$ bands. Considering the blazar sample, all FSRQs were variable at greater than 99\% confidence level in 
all three filters. Similar results were obtained for all except two BL Lac objects. The source J0127$-$0821 is variable in $g$ and $r$ bands, though $i$ band data are unavailable. The source J1735$+$5650 is variable in $g$ band, a probable variable in the $r$ band while non-variable in the $i$ band. 

\subsubsection{Amplitude of flux variations}
We calculated the fractional variability amplitude ($F_{var}$) for the light curves to get an idea about the amplitude of flux variations \citep{vaughan2003, sesar2007}. It is defined as : 
\begin{equation}
F_{\text{var}} = \sqrt{\frac{S^2 - \bar{f}_{\text{err},i}^2}{\bar{f_i}^2}},
\end{equation}

where $S$ is the sample variance of the light curve, $\bar{f_i}$ is the mean of the $f_i$ measurements, and $\bar{f}^2_{\text{err}}$ is the mean square error of each individual error $f_{\text{err},i}$, given by 
\begin{equation}
S^2 = \dfrac{1}{n-1} \sum_{i=1}^{n} (f_i - \bar{f_i})^2,
\end{equation}

\begin{equation}
\bar{f}_{\text{err},i}^2 = \dfrac{1}{n} \sum_{i=1}^{n} (f_{\text{err},i}^2).
\end{equation}

The uncertainty in F$_{var}$ is defined as \citep{vaughan2003}
\begin{equation}
\small
F_{var,err} = \sqrt{\left( \sqrt{\dfrac{1}{2n}}\dfrac{\bar{f}^2_{err}} {\bar{f_i}^2 F_{var}}\right)^2 +  \left(\sqrt{\dfrac{\bar{f}^2_{err}}{n}} \dfrac{1}{\bar{f_i}}\right)^2}.
\end{equation}

\begin{figure}[h!]
\centering
\vbox{
     \includegraphics[width=0.98\columnwidth]{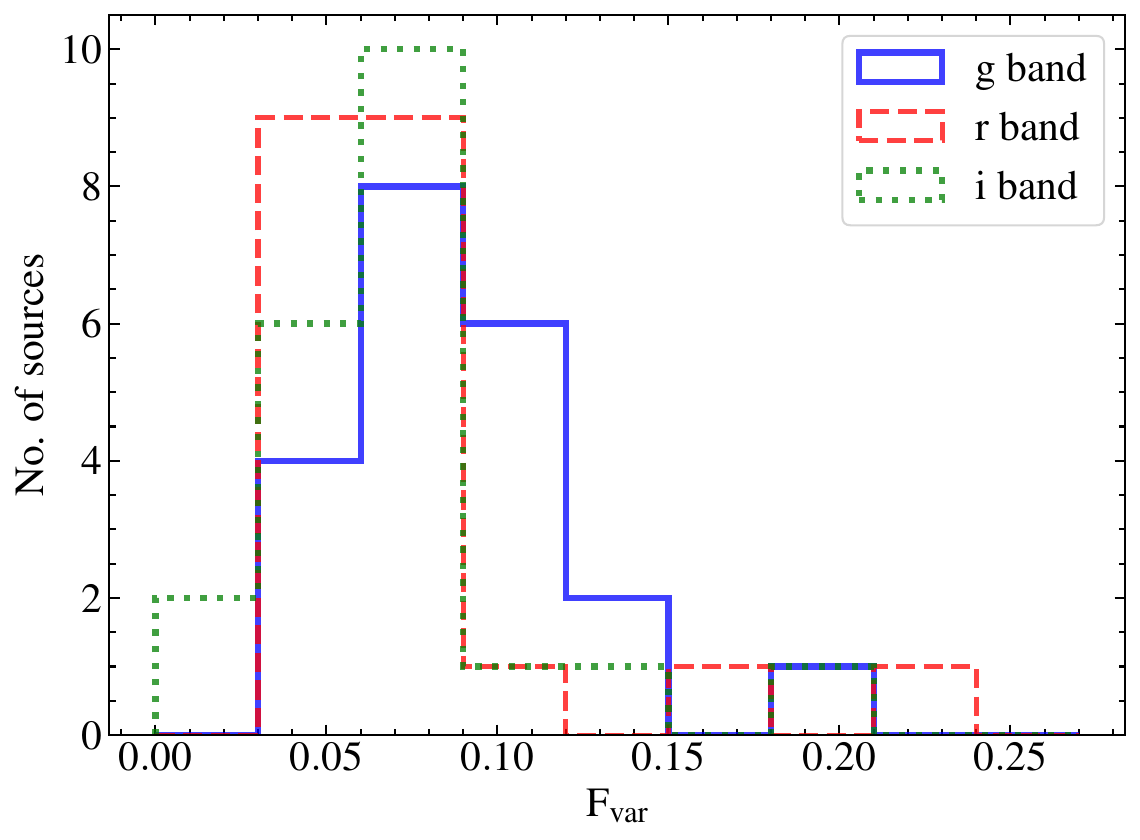}
     \includegraphics[width=\columnwidth]{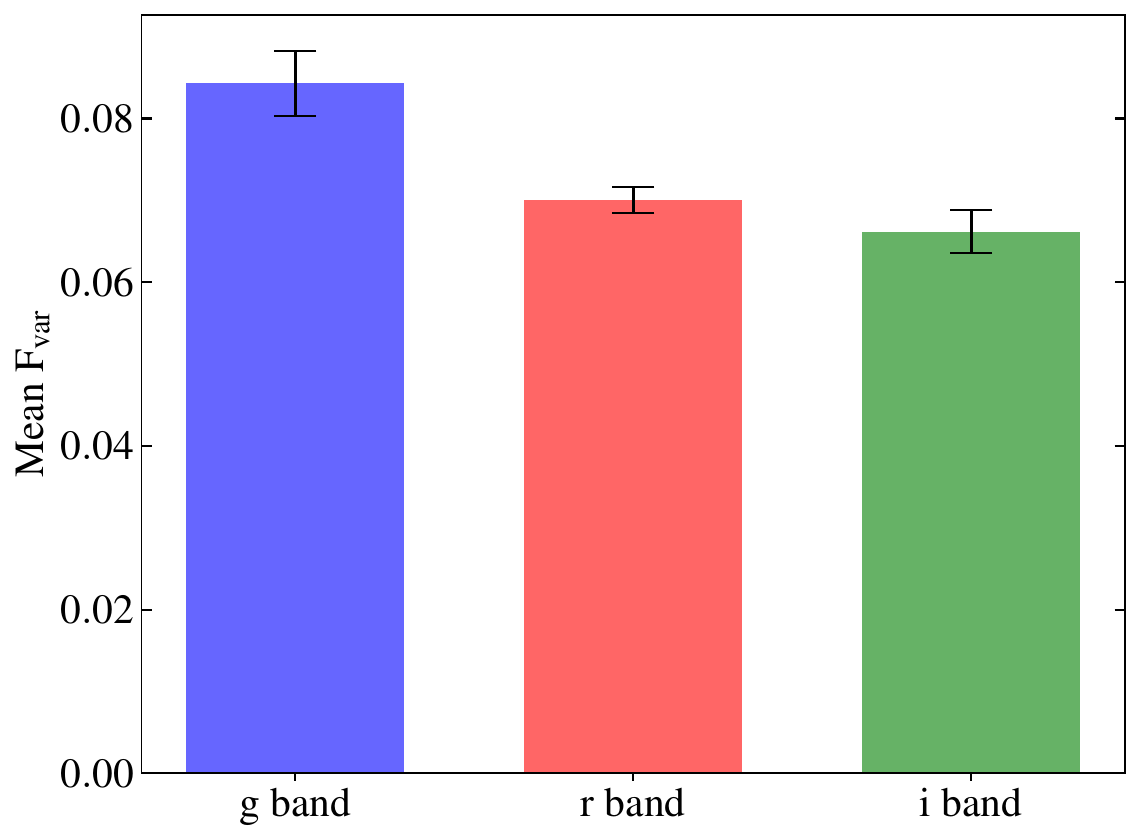}
     }
\caption{Distributions of the fractional variability amplitude (F$_{var}$) in $g$, $r$ and $i$ bands (top panel) and the mean F$_{var}$ in $g$, $r$ and $i$ bands (bottom panel) for CSOs.}
\label{figure-2}
\end{figure}

The derived $F_{var}$ values are tabulated for all CSOs in Table \ref{table2}. The $F_{var}$ values for 2, 3, and 1 CSOs could not be calculated in the $g$, $r$, and $i$ bands, respectively, due to large uncertainties in the measurements. Out of 27 objects for which $F_{var}$ could be estimated in all three filters, we further selected those 21 CSOs that have exhibited significant variability in all three bands. For these objects, the distributions of $F_{var}$ values are presented in the top panel of Figure \ref{figure-2} and mean $F_{var}$ values are shown in the bottom panel of Figure \ref{figure-2}. 
CSOs are found to show more variability in the $g$ band than the $r$ and $i$ bands, while they show similar variability in the $r$ and $i$ bands. We also looked for similarities and/or differences in the fractional amplitude of flux variation between the CSOs that were identified as galaxies and quasars based on their absolute $B$-band magnitudes. A moderate difference was noticed in the variability behaviour between CSOs that are galaxies and quasars in $g$, $r$ and $i$ bands with the latter appearing to be more variable, the difference being most significant in the $r$ band (Table \ref{table-3}). The prominence of the radio core in CSSs/CSOs suggests the quasars to be inclined at smaller angles to the line of sight than galaxies \citep[e.g.][]{Saikia1995MNRAS.276.1215S,ODea2021}. However, they are both inclined at much larger angles to the line of sight than blazars, and the difference in orientation between CSO quasars and galaxies 
need to be probed using a larger sample as there are only 4 quasars in this sample (Table \ref{table-3}). 

\begin{figure}[h!]
\centering
	\includegraphics[width=6.15cm,height=4.8cm]{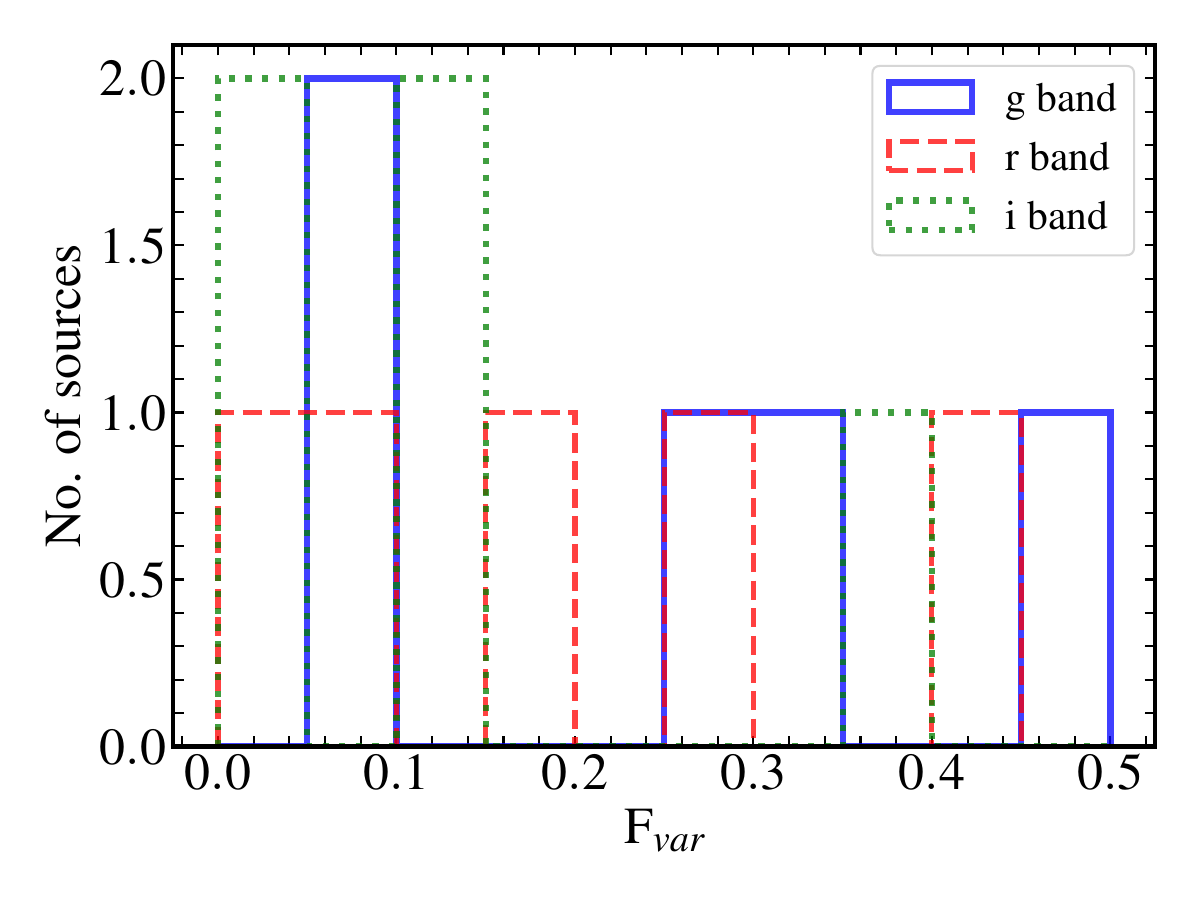}
	\includegraphics[width=6.15cm,height=4.8cm]{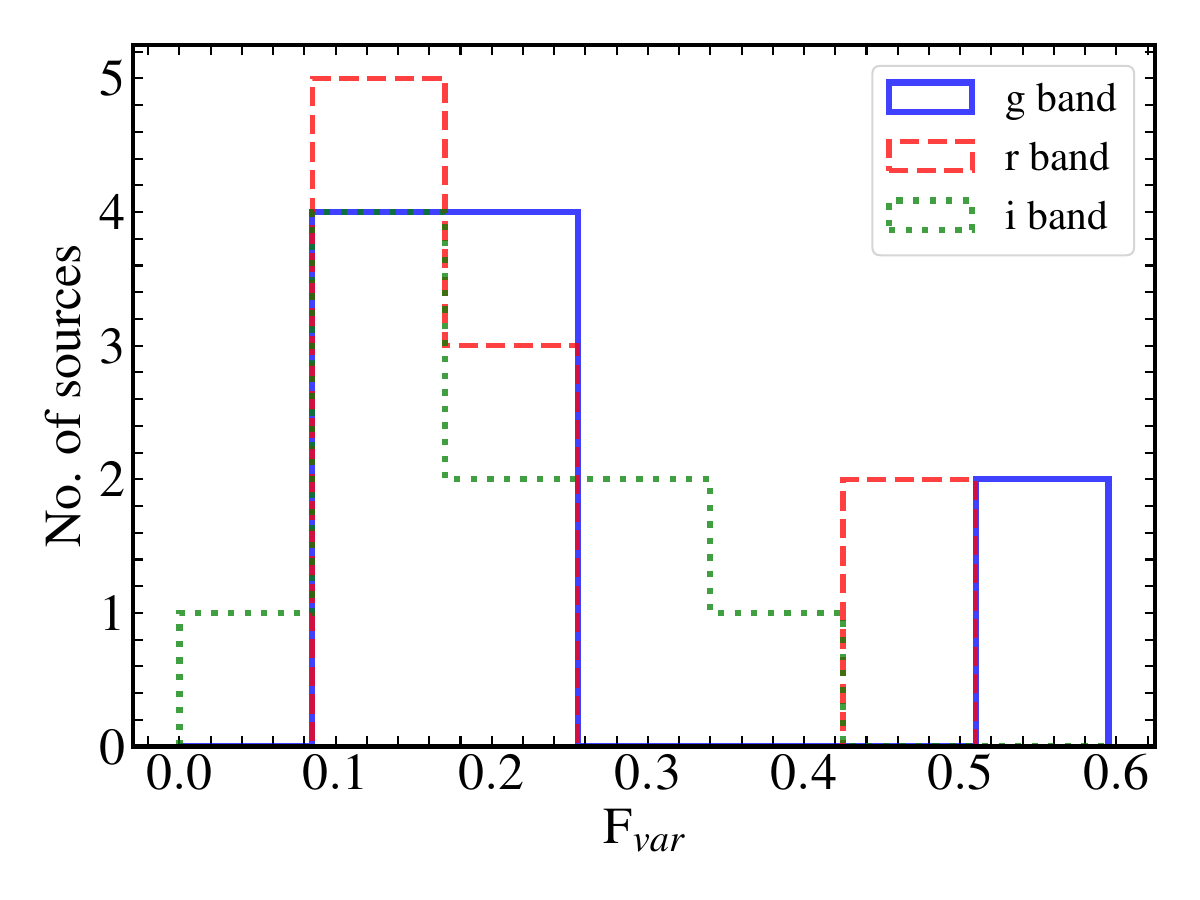}
    \caption{Distributions of the fractional variability amplitude (F$_{var}$) in $g$, $r$ and $i$ bands
for FSRQs (top panel) and BL Lacs (bottom panel).}
    \label{figure-4}
\end{figure}

\begin{figure}[h!]
\centering
	\includegraphics[width=\columnwidth,height=5.5cm]{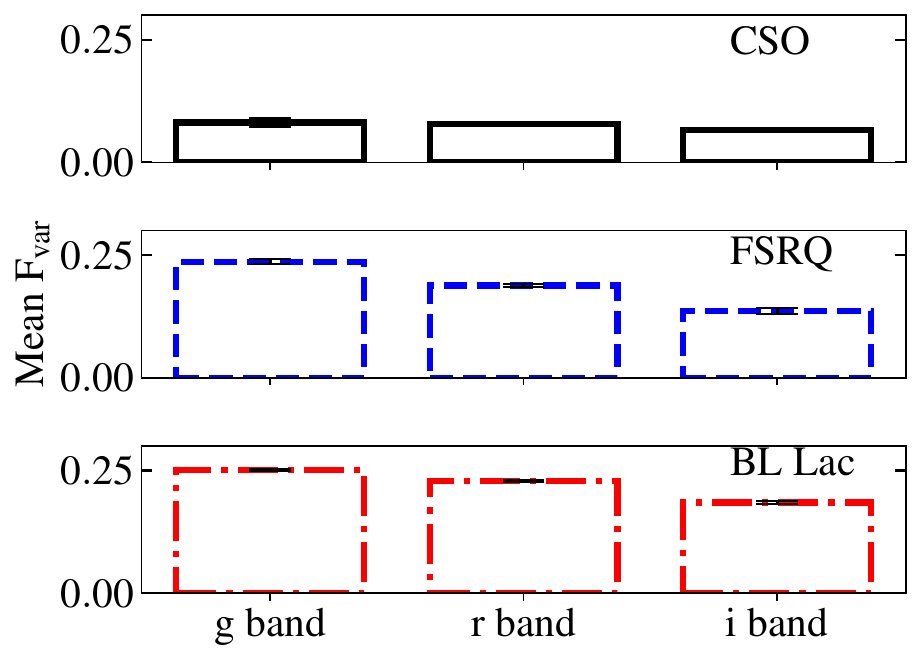}
    \caption{The mean fractional amplitude of variability for the matched sub-sample of CSOs, FSRQs, and BL Lac sources in $g$, $r$ and $i$ bands.}
    \label{figure-5}
\end{figure}

Similar to CSOs, for blazars too, we calculated the $F_{var}$ in all three bands (Table \ref{table2}). The distributions of $F_{var}$ values in different filters for both FSRQs and BL Lacs are plotted in Figure \ref{figure-4}. Further, we selected the comparison sub-sample of CSOs  and blazars 
that have exhibited significant variability in all three bands. A comparison
of the $F_{var}$ values for this matched sample are given in Table \ref{table-5} and shown in Figure \ref{figure-5}.  We found that blazars exhibit significant flux variations, with the mean fractional variability amplitude being approximately three times higher in the $g$ band and two times higher in the $r$ and $i$ bands compared to CSOs. Also, blazars exhibited larger variability in the shorter-wavelength $g$ band compared to the $r$ band, and even less in the i-band, following the trend $F_{var,g}$ $>$ $F_{var,r}$ $>$ $F_{var,i}$. 

\subsection{Spectral Variability}
\begin{figure*}
\centering
    \includegraphics[width=3cm]{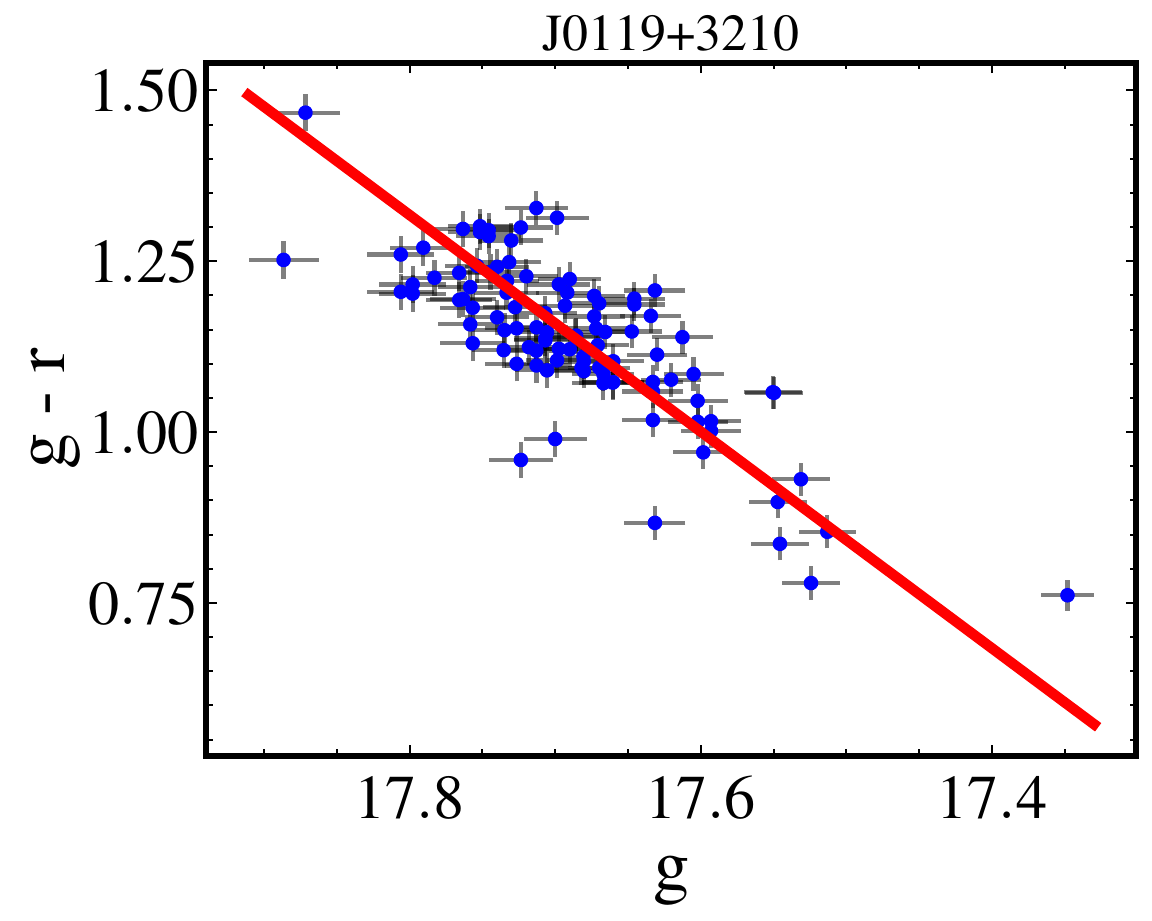}
    \includegraphics[width=3cm]{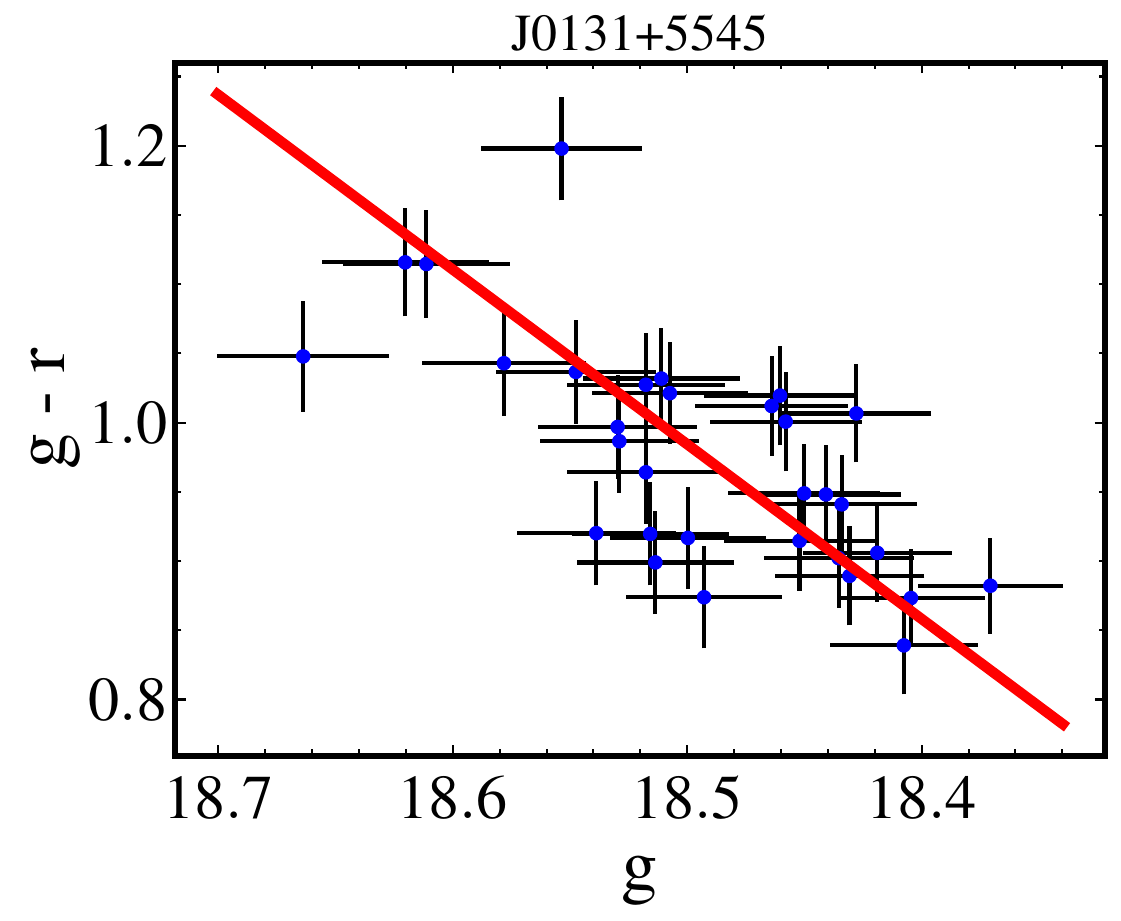}
    \includegraphics[width=3cm]{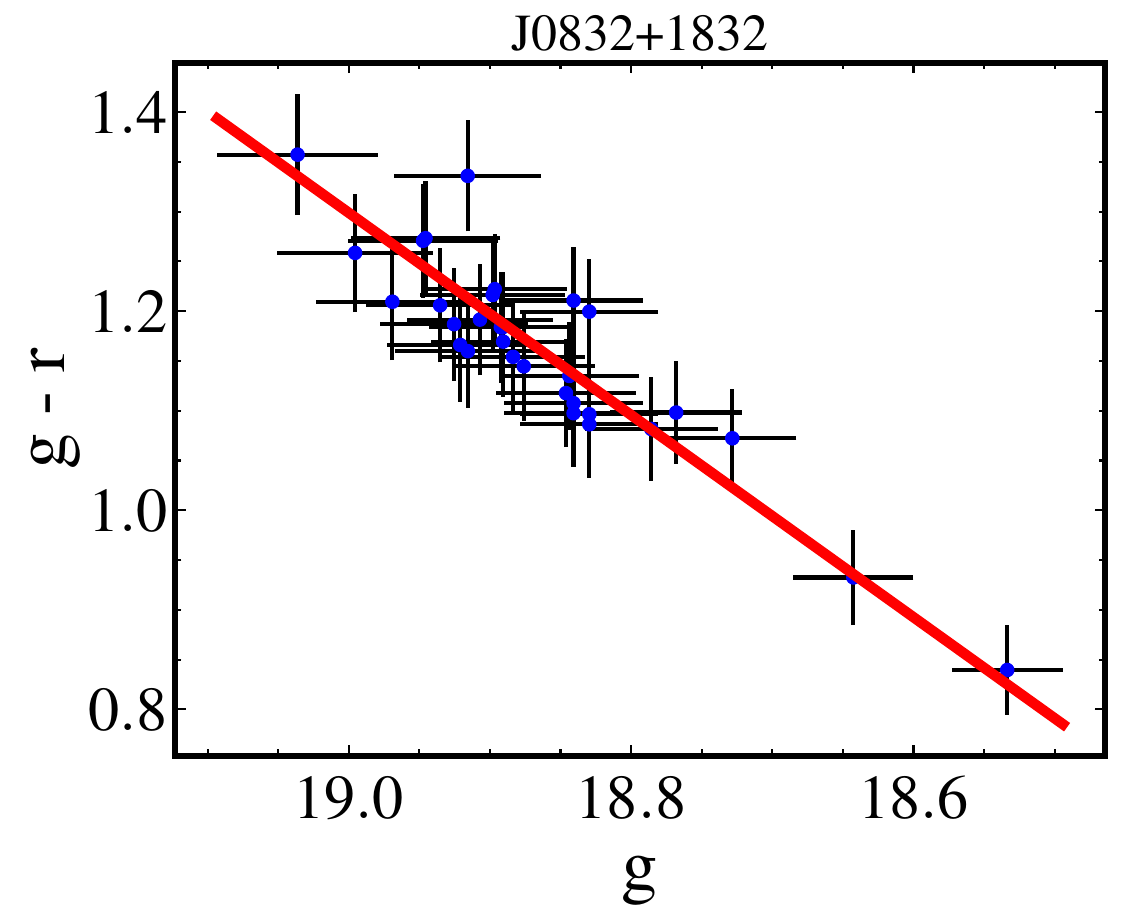}
    \includegraphics[width=3cm]{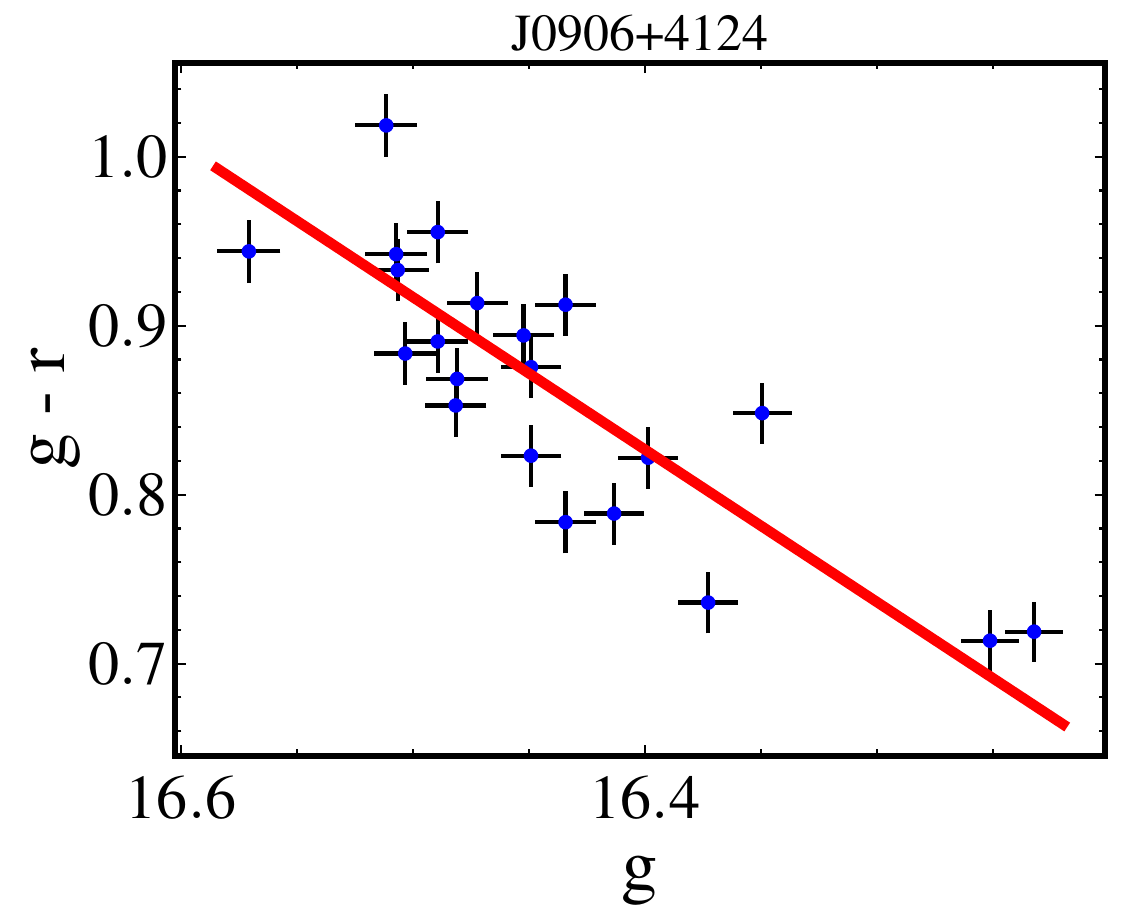}
    \includegraphics[width=3cm]{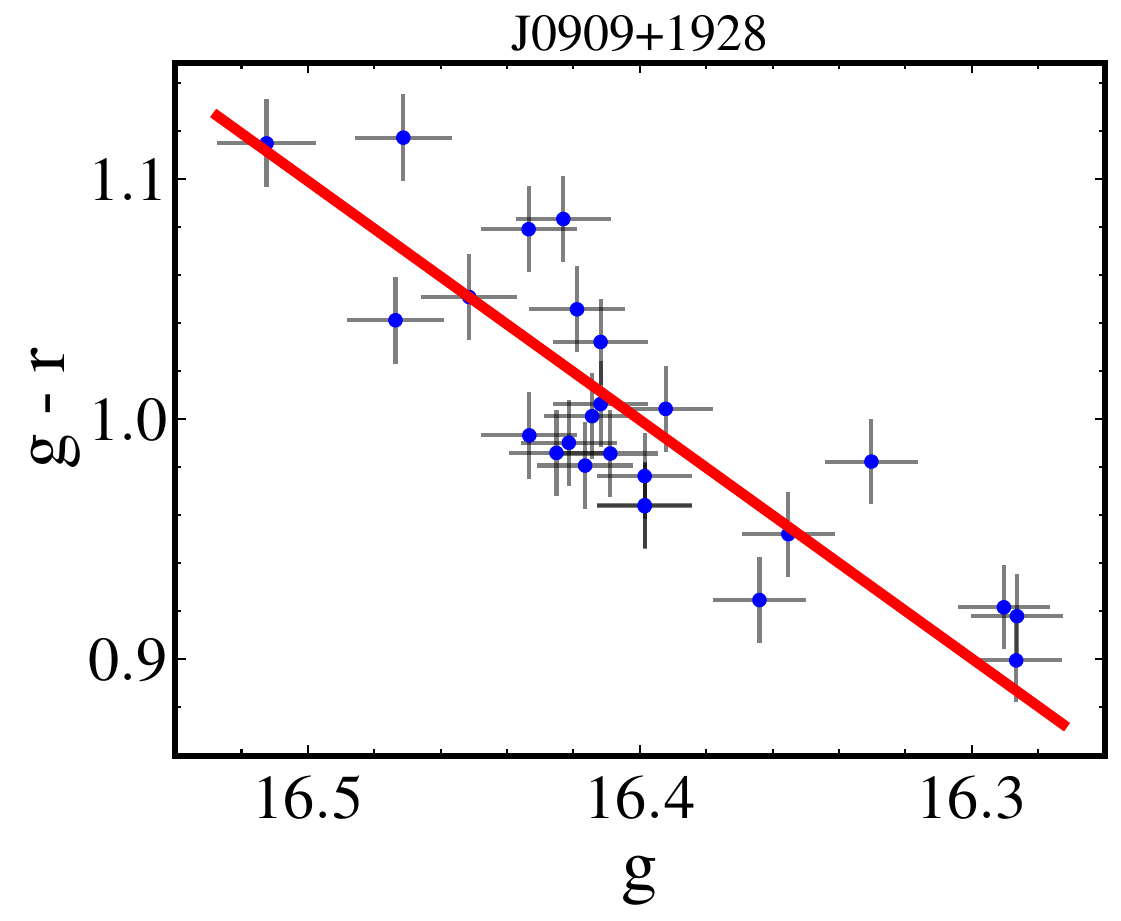}
    \includegraphics[width=3cm]{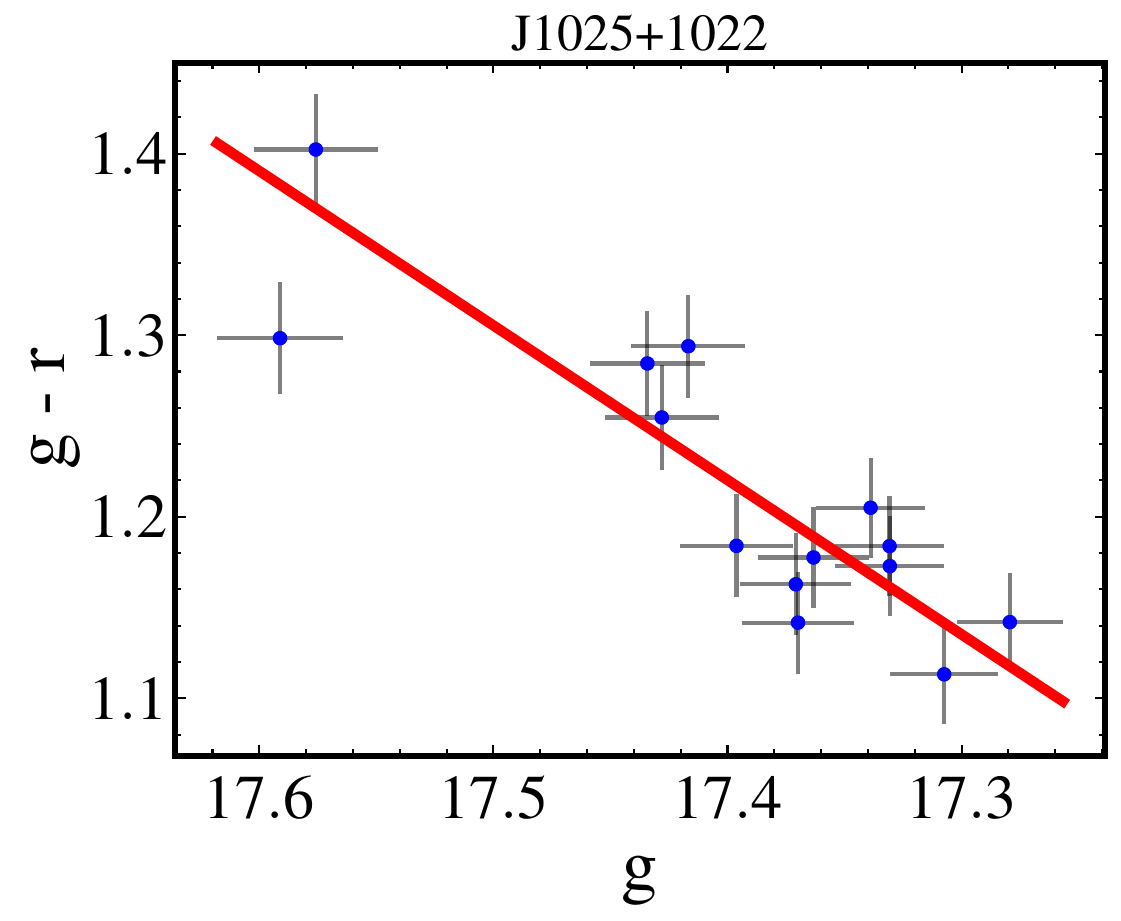}
    \includegraphics[width=3cm]{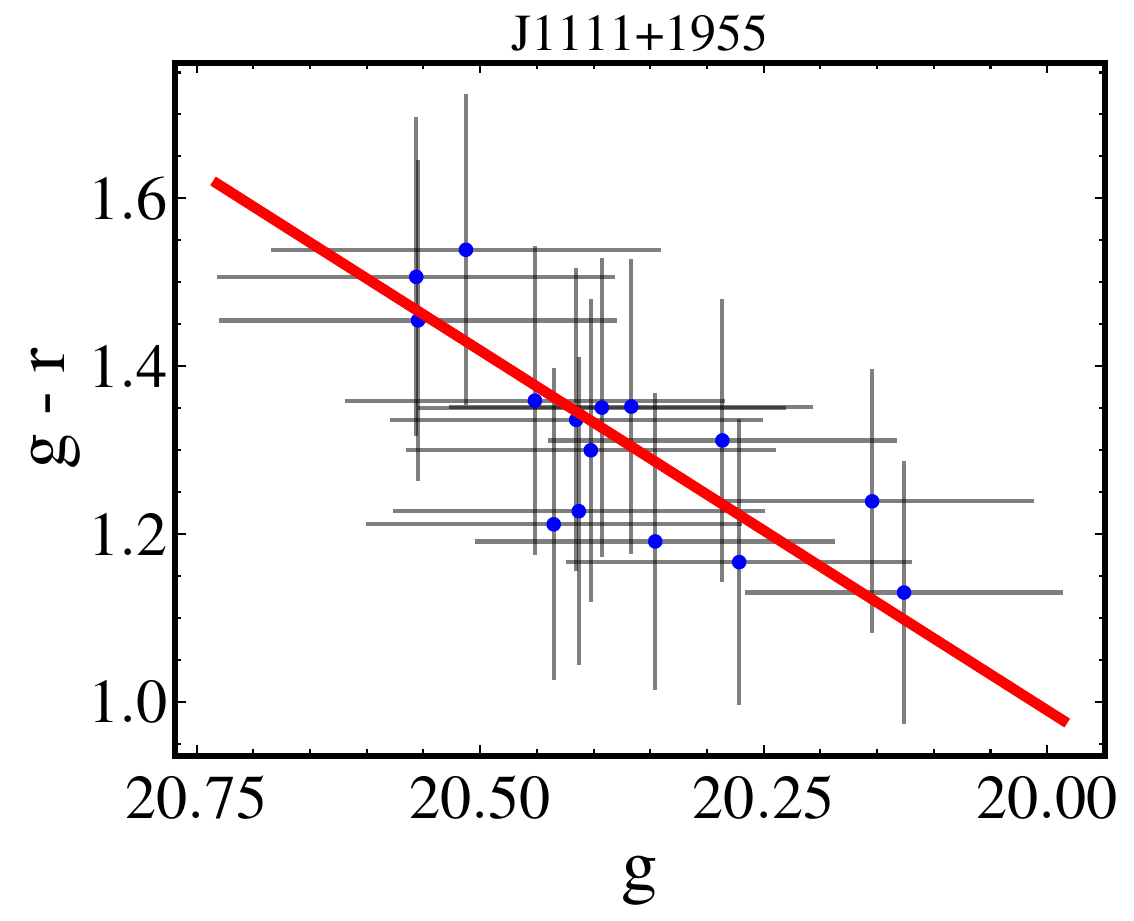}
    \includegraphics[width=3cm]{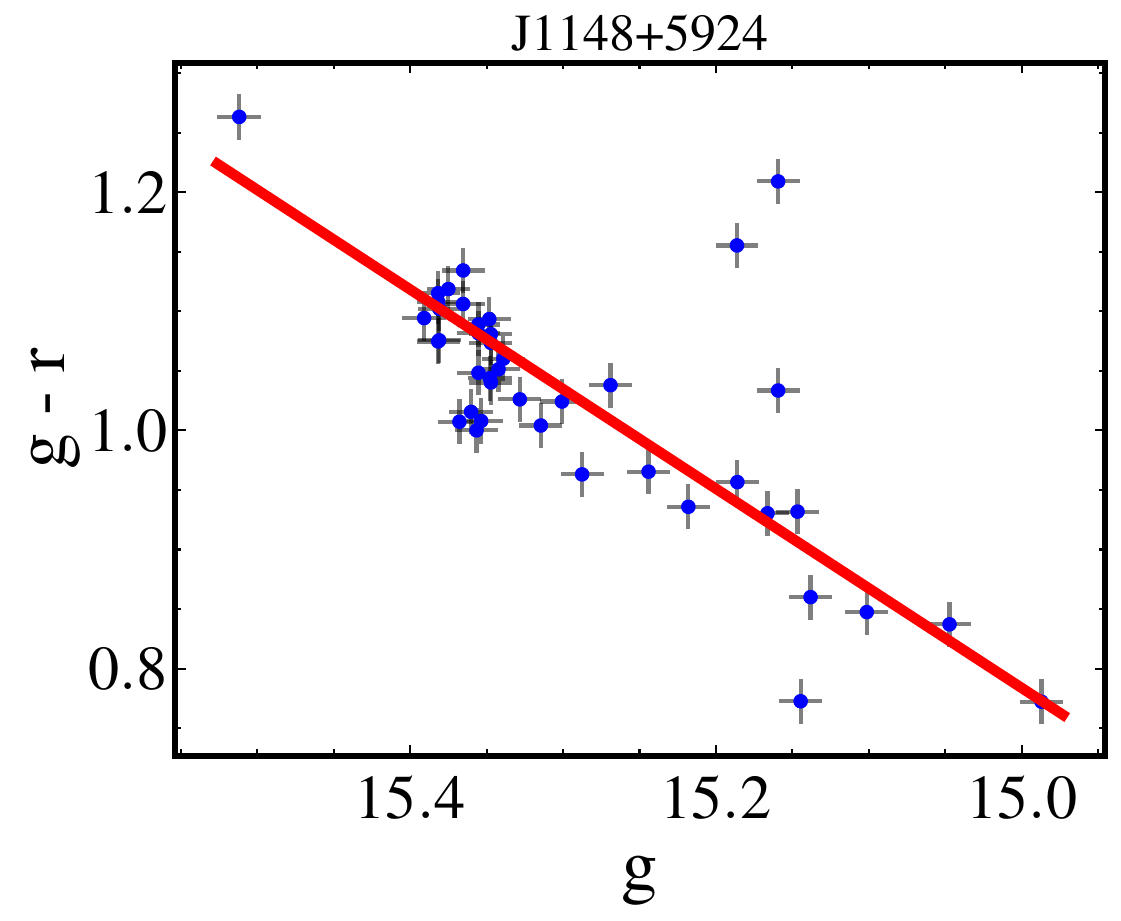}
    \includegraphics[width=3cm]{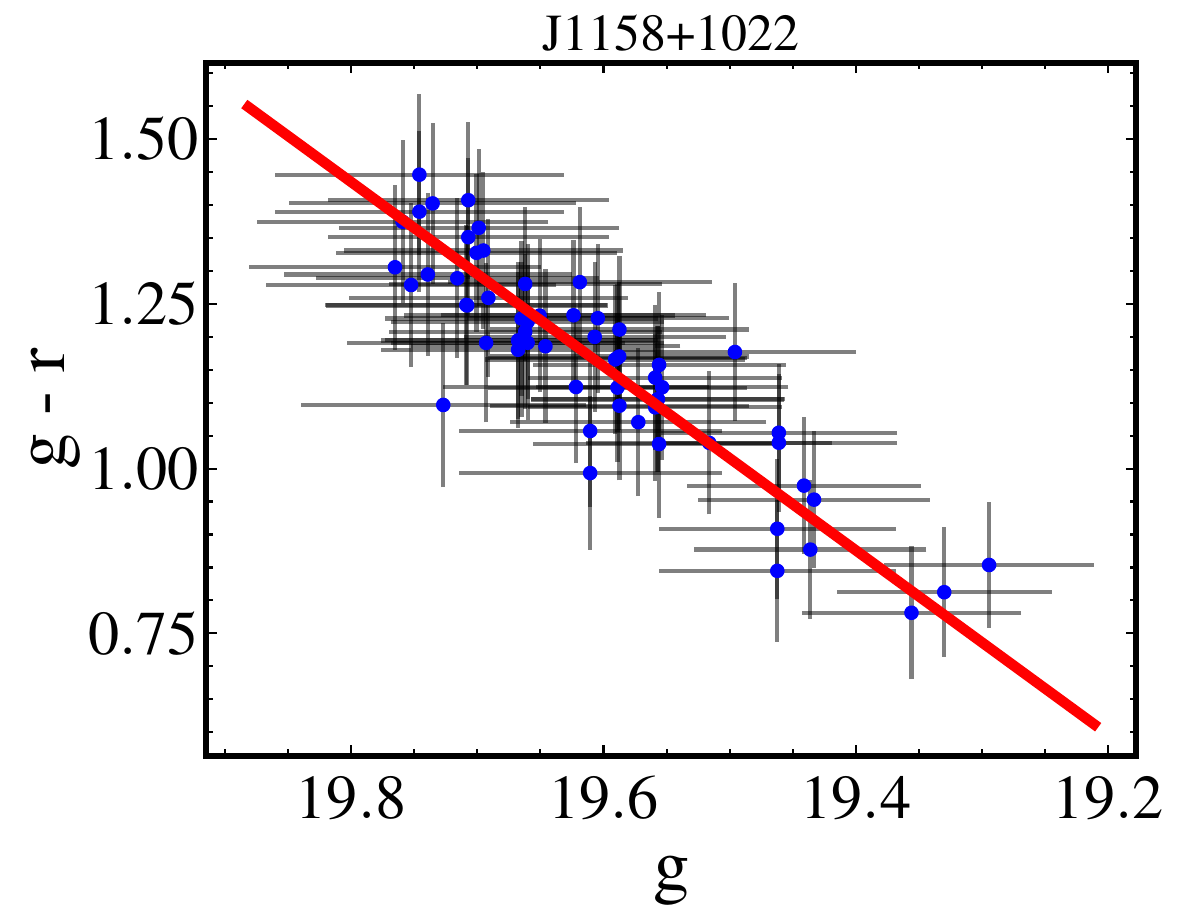}
    \includegraphics[width=3cm]{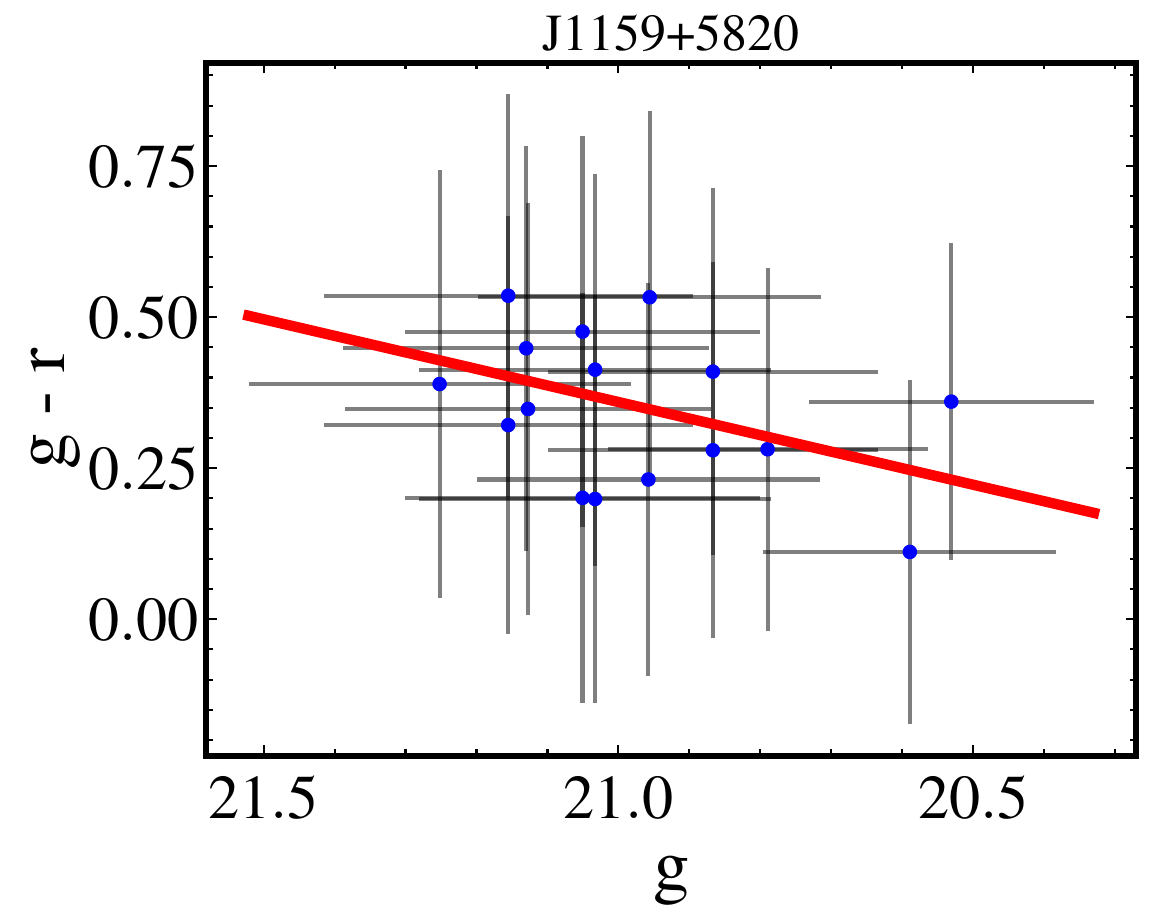}
    \includegraphics[width=3cm]{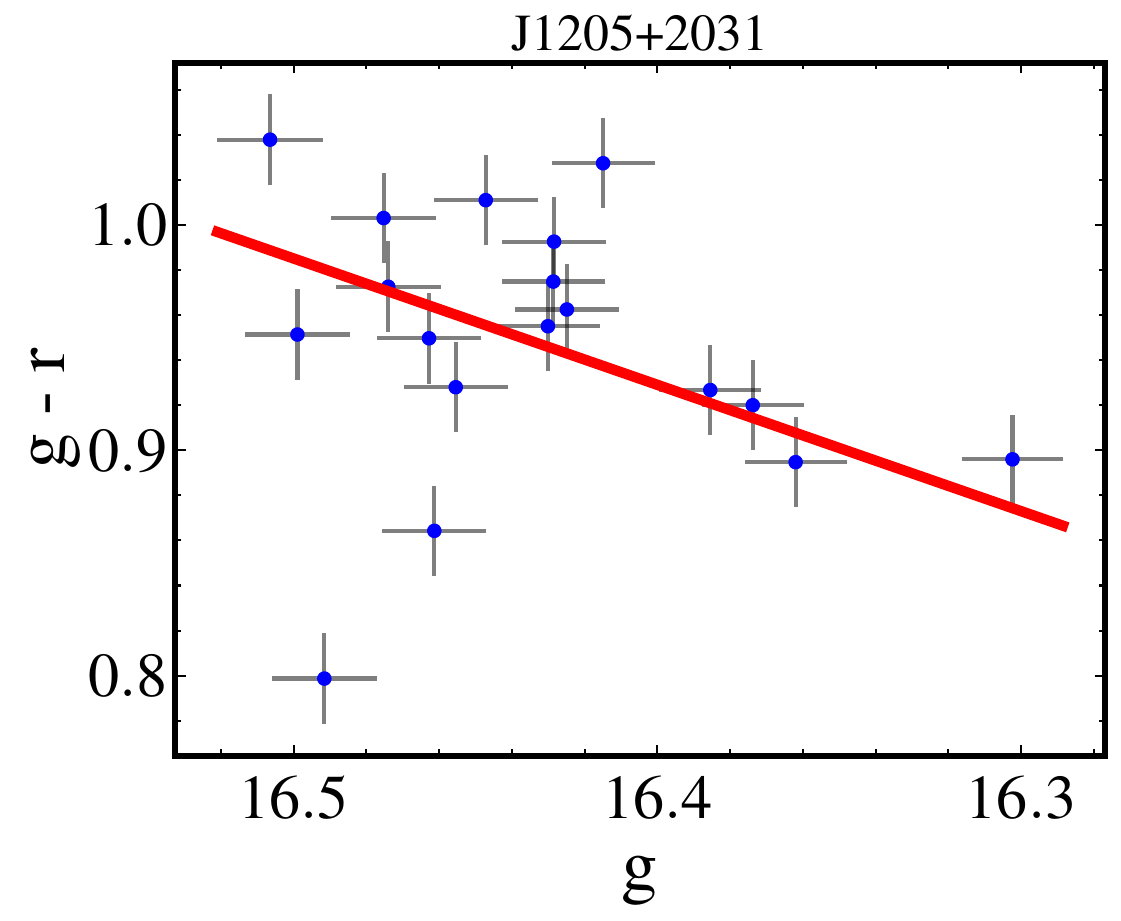}
    \includegraphics[width=3cm]{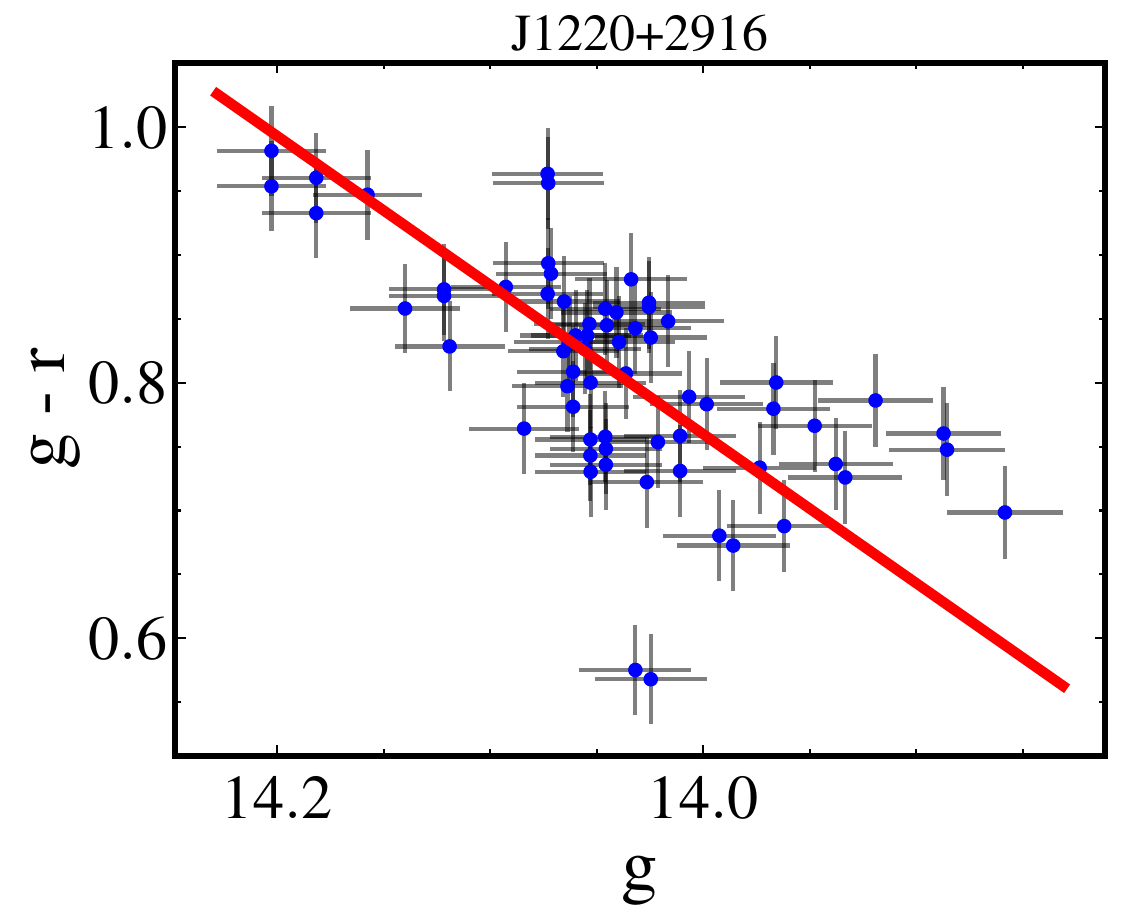}
    \includegraphics[width=3cm]{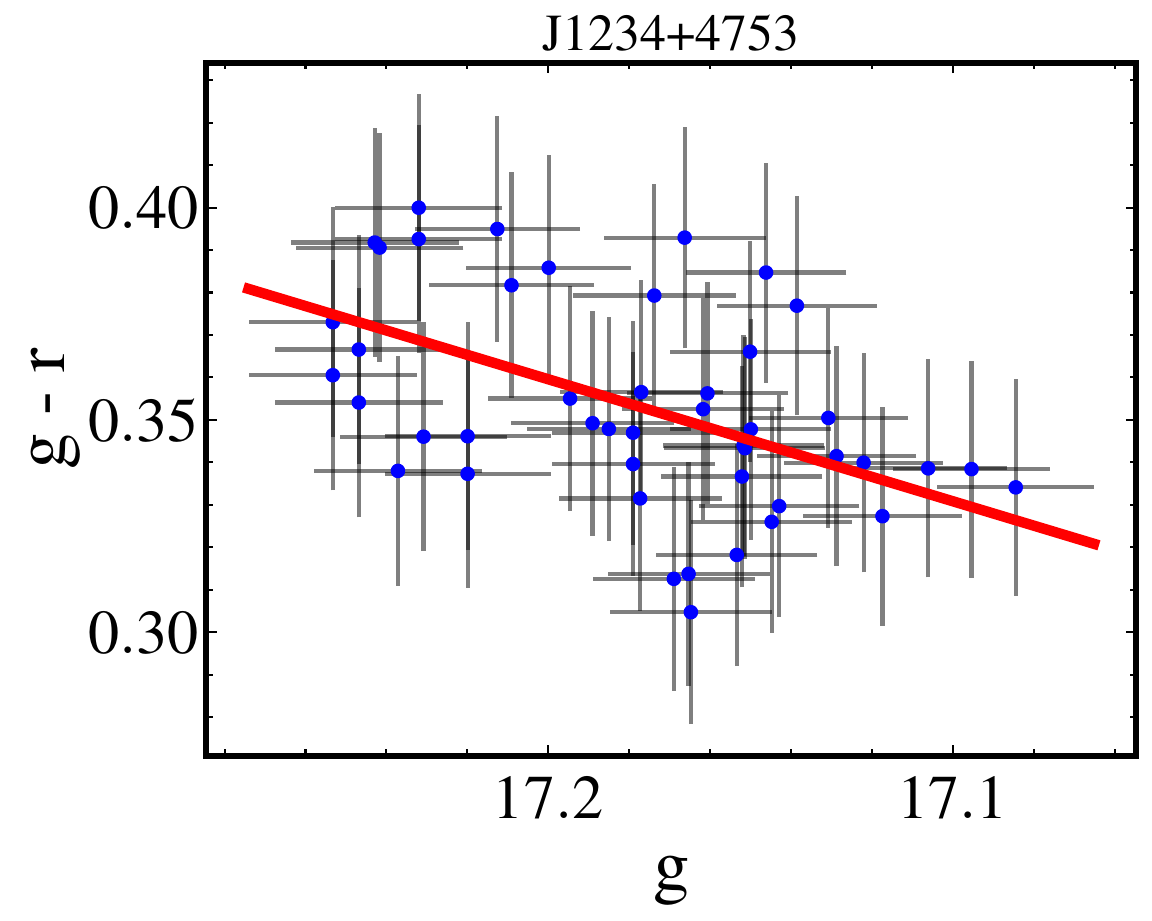}
    \includegraphics[width=3cm]{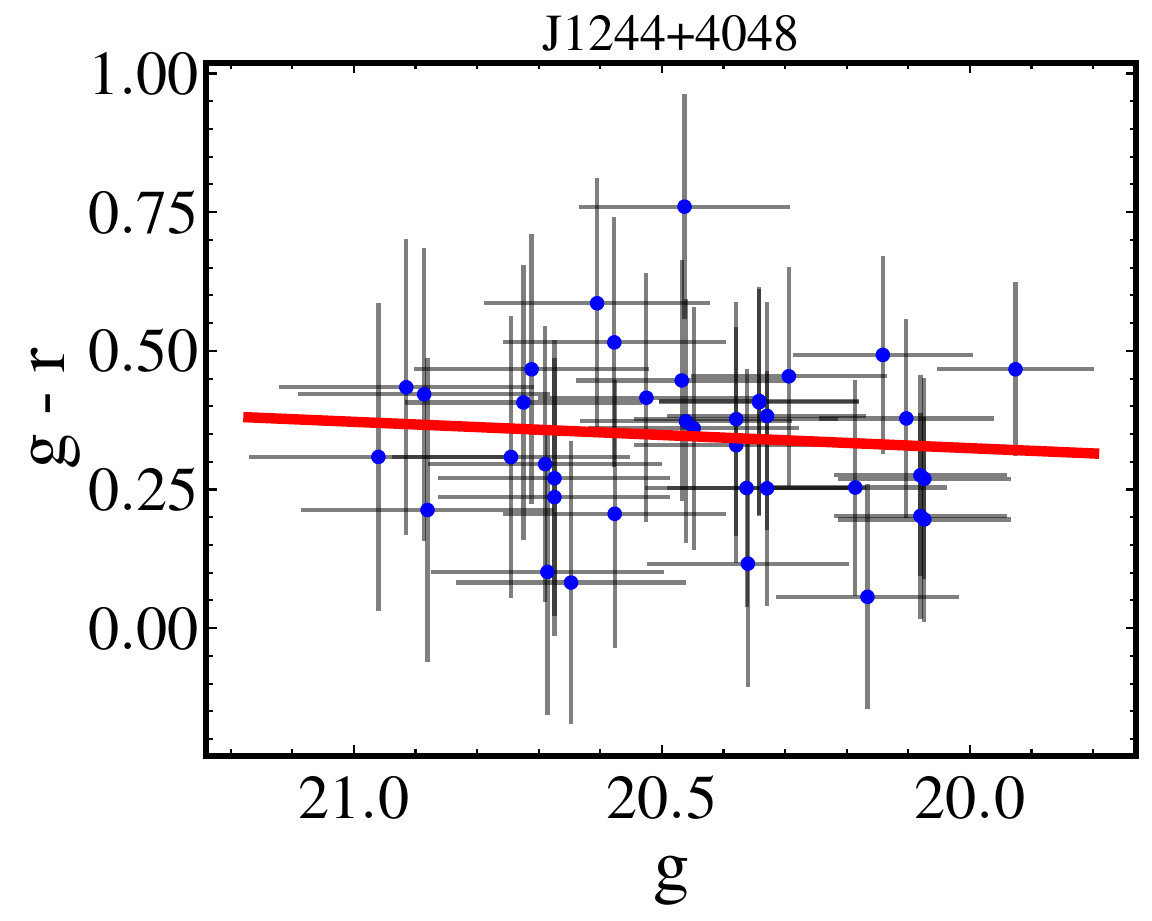}
    \includegraphics[width=3cm]{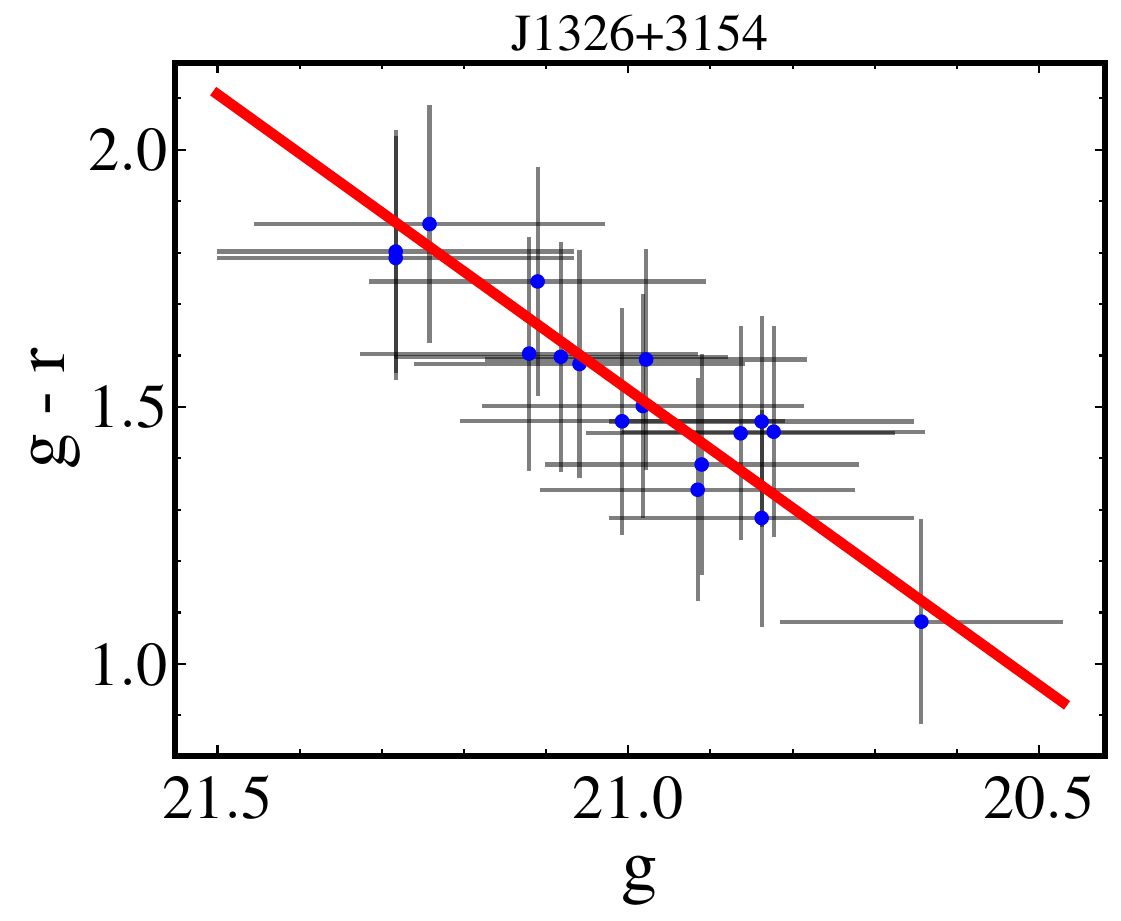}
    \includegraphics[width=3cm]{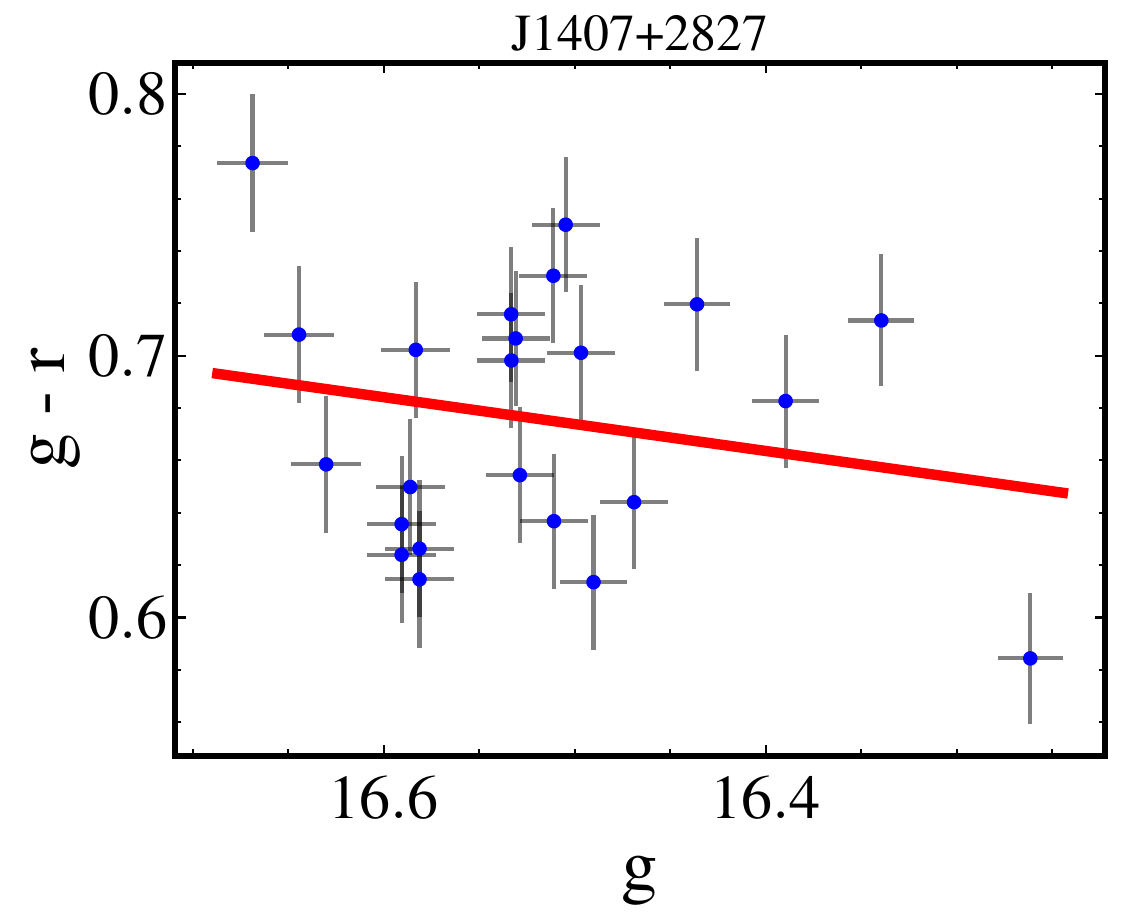}
    \includegraphics[width=3cm]{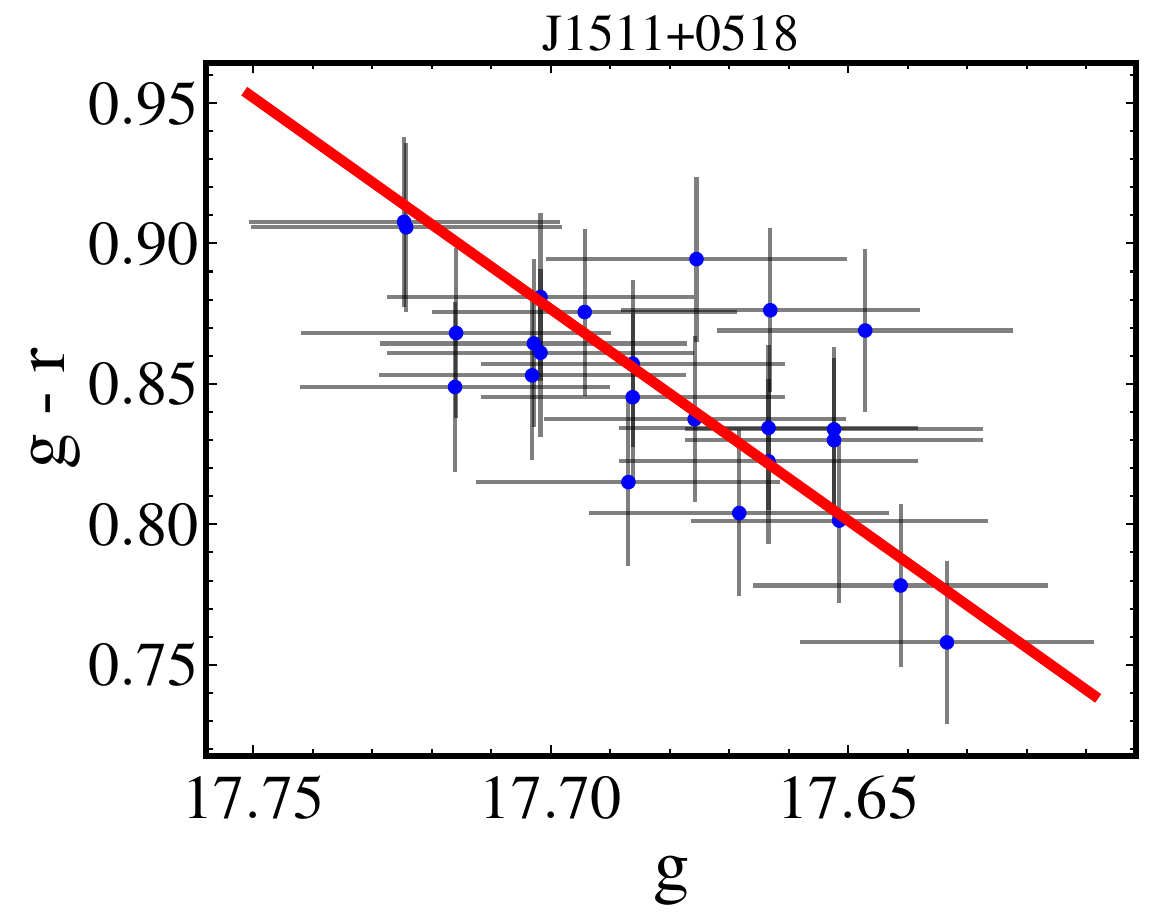}
    \includegraphics[width=3cm]{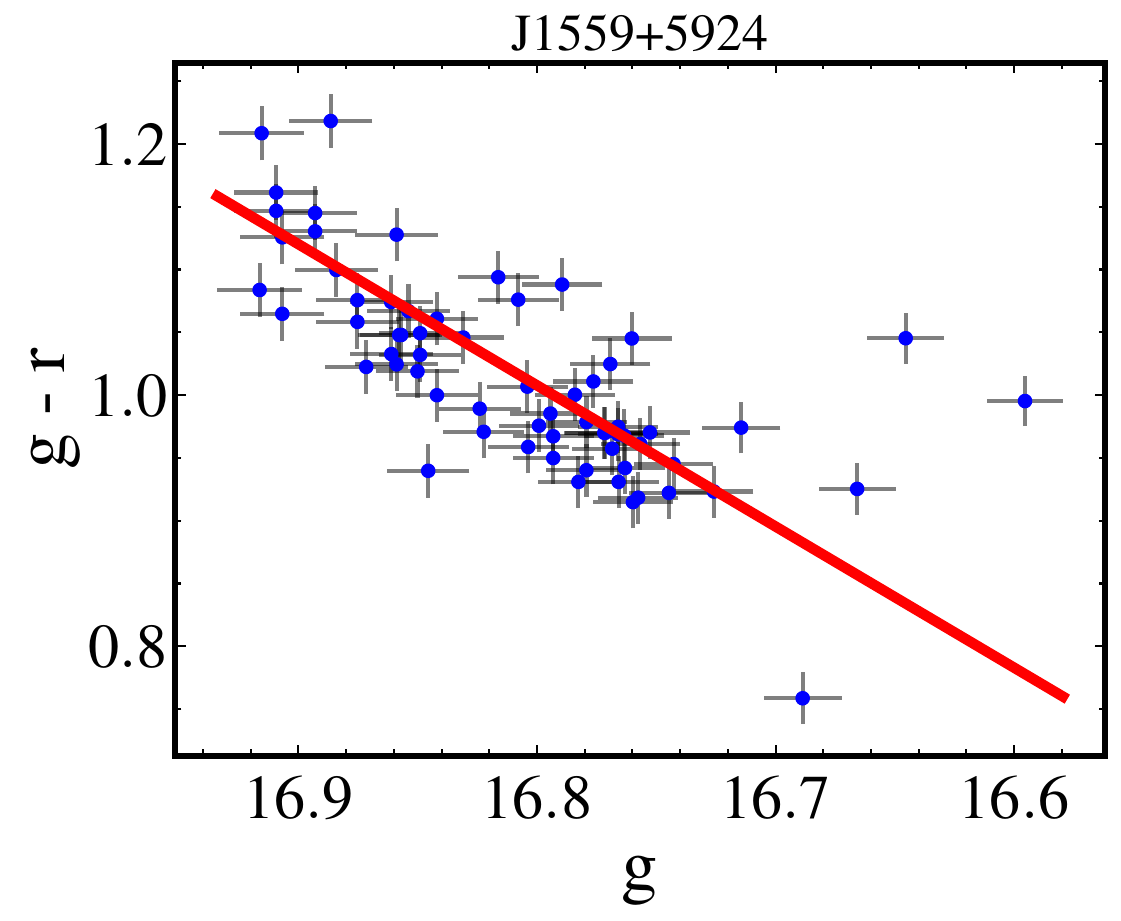}
    \includegraphics[width=3cm]{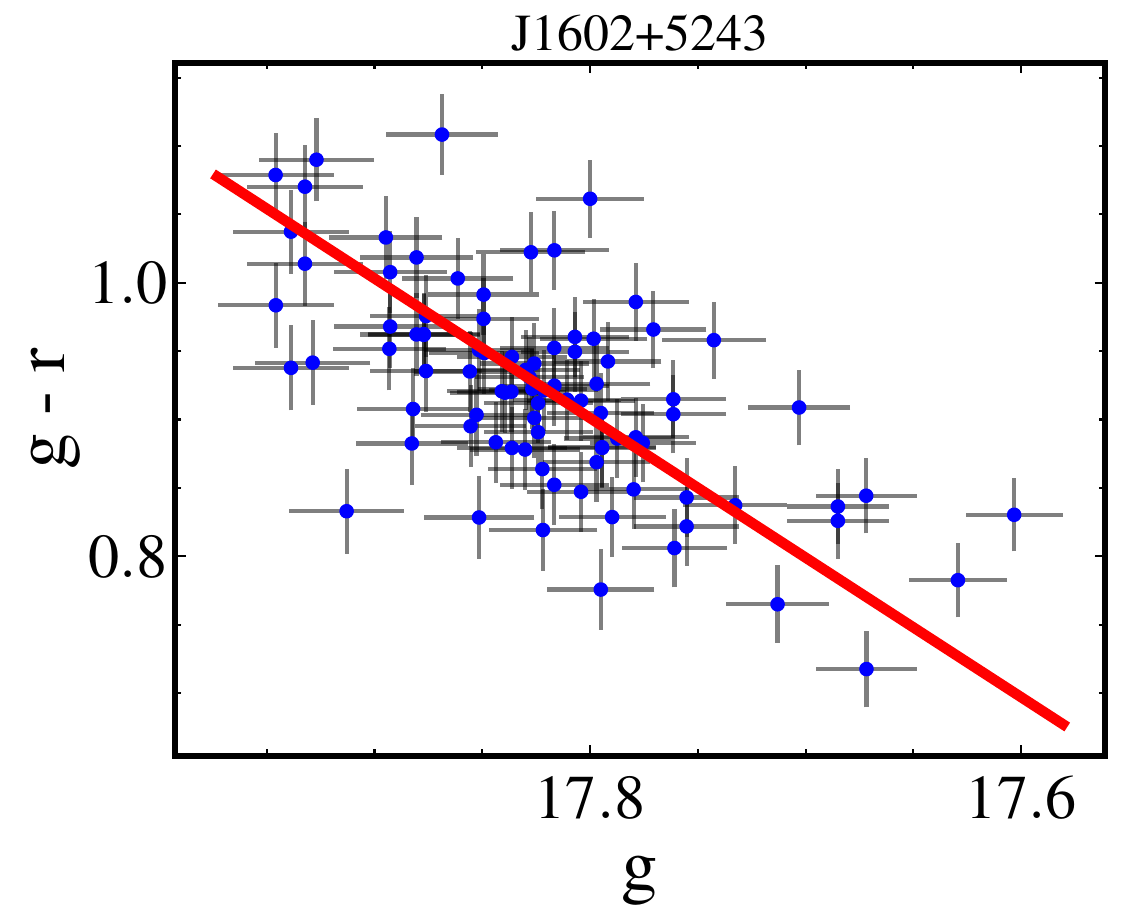}
    \includegraphics[width=3cm]{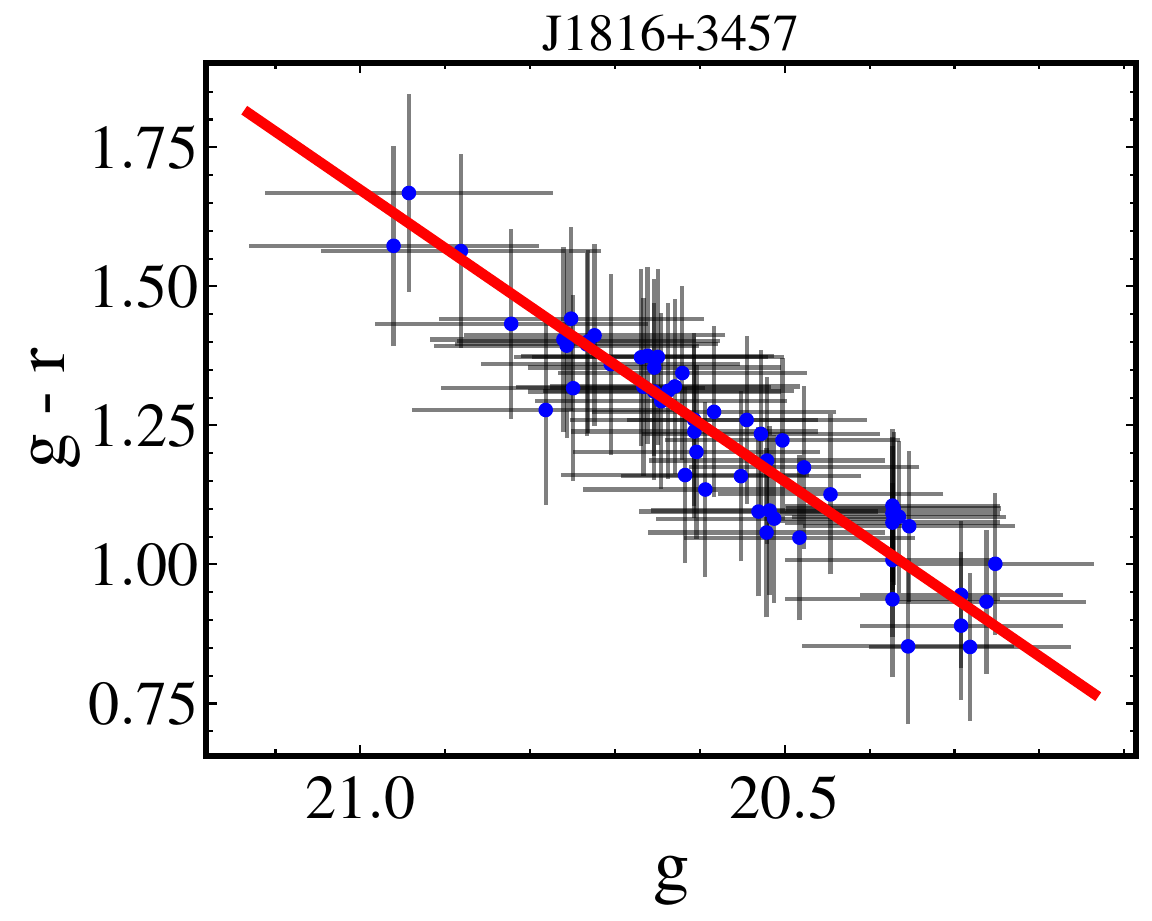}
    \includegraphics[width=3cm]{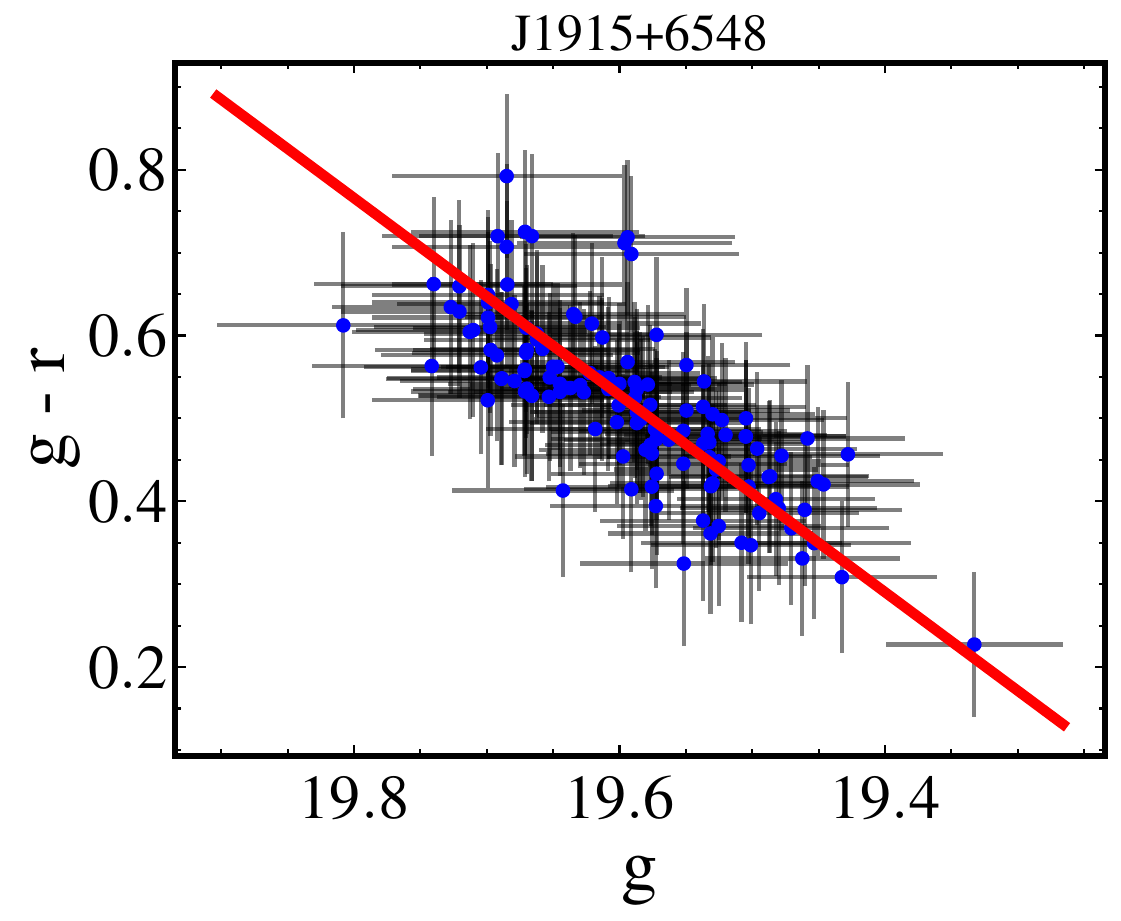}
    \includegraphics[width=3cm]{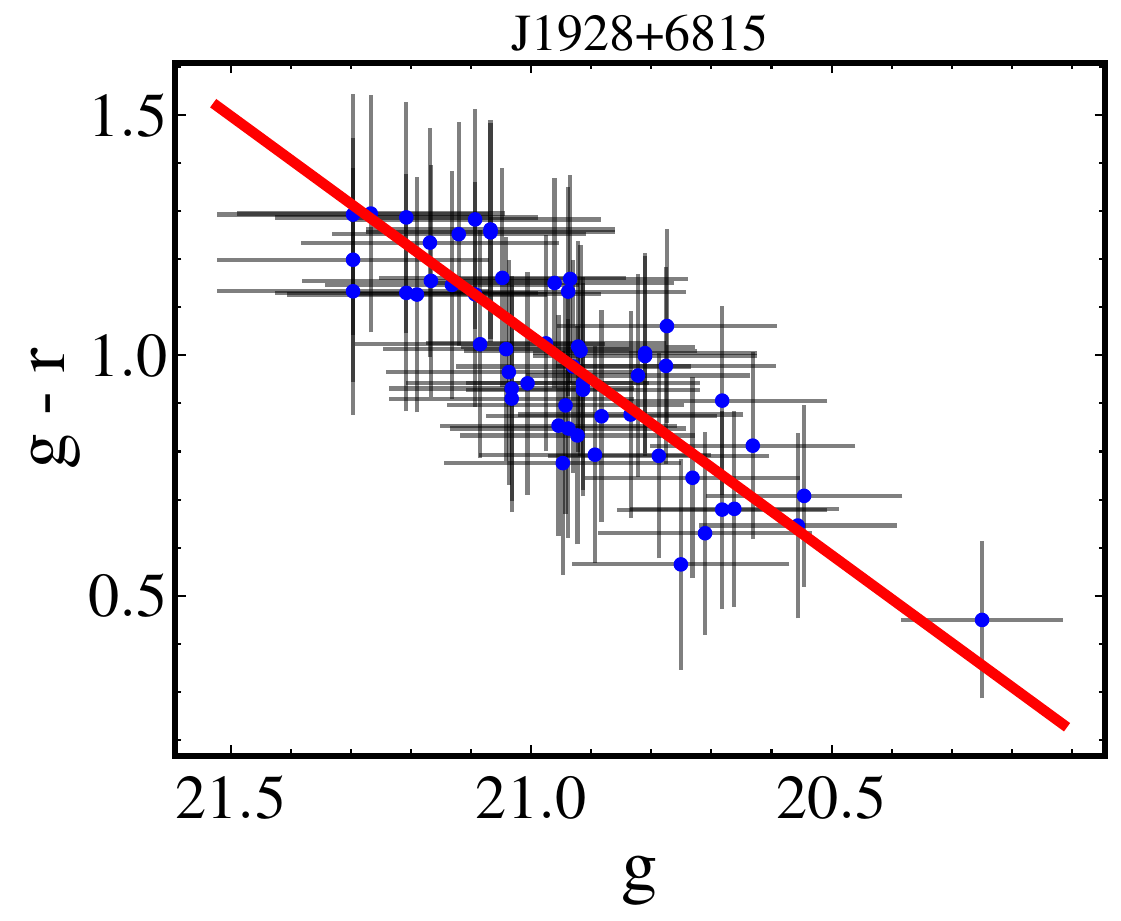}
    \includegraphics[width=3cm]{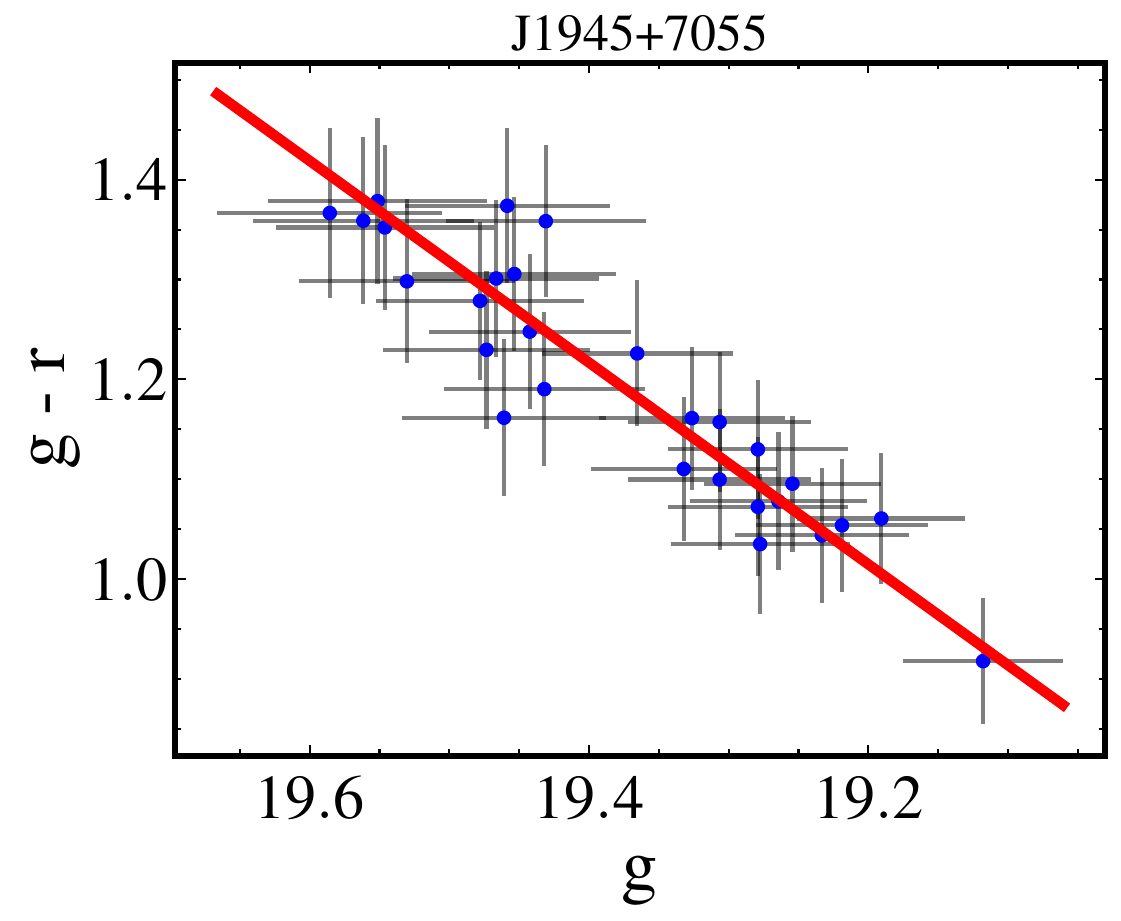}
    \includegraphics[width=3cm]{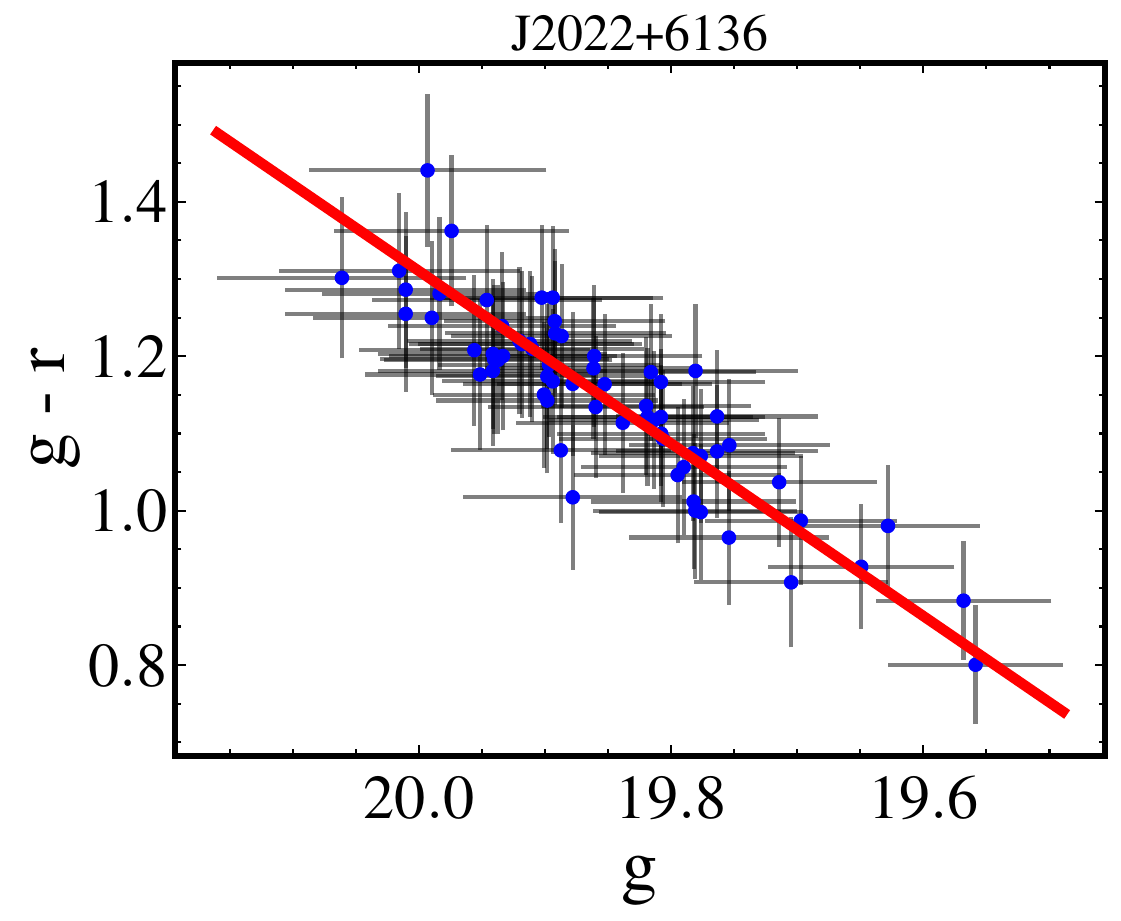}
    \includegraphics[width=3cm]{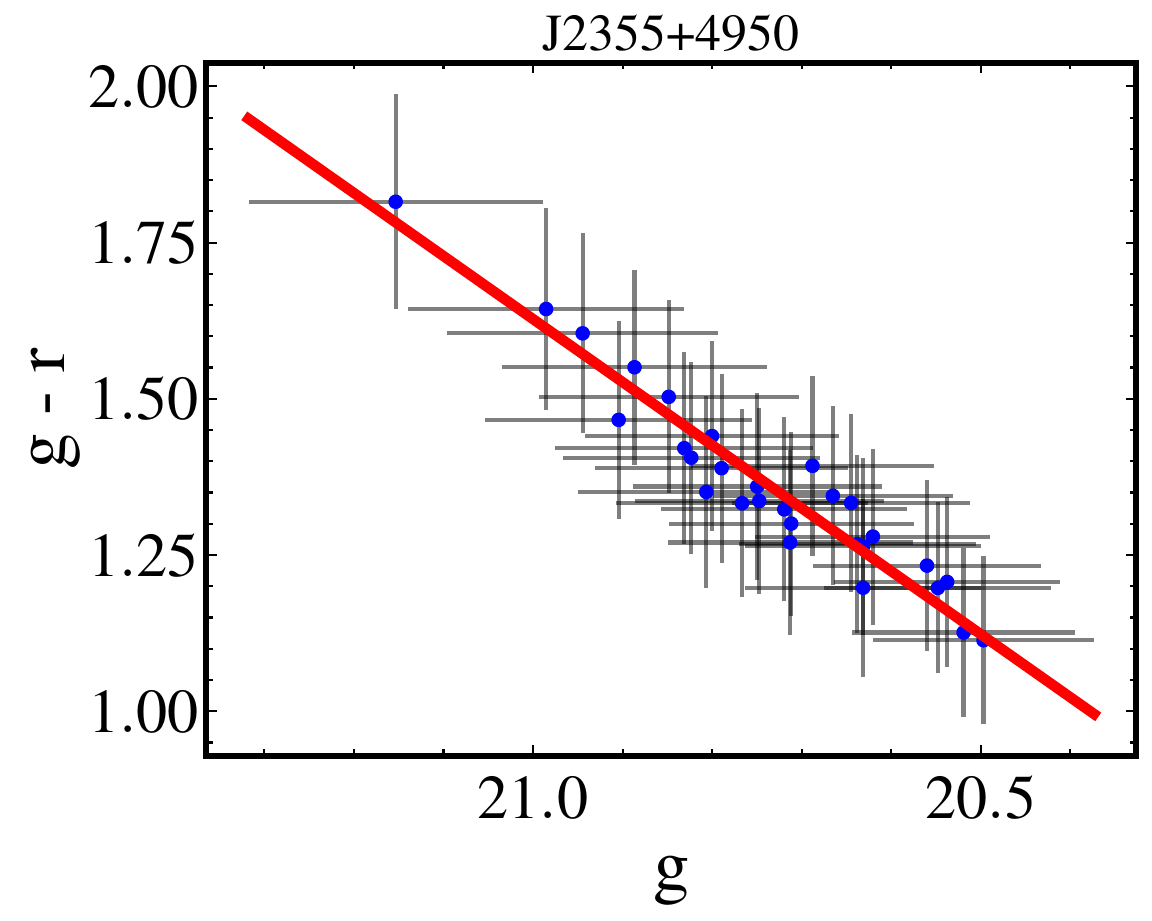}
    \caption{Colour ($g$$-$$r$) against magnitude ($g$-band) diagram for CSOs. The solid line is the weighted linear least squares fit to the data. }
    \label{figure-3}
\end{figure*}

To investigate possible spectral changes over longer timescales, we generated
colour-magnitude diagram, wherein we looked for variations in the $g$$-$$r$ colour
against the $g$-band brightness. This requires simultaneous observations in g- and r-band filters, and therefore, we considered those $g$- and $r$-band measurements 
that were taken within 30 minutes of each other. We imposed an additional constraint
on the sources to have more than five colour measurements. This led us to a sample of 25 CSOs for analysis of colour variability. The color-magnitude diagrams for 25
CSOs are shown in Figure \ref{figure-3}. Also shown in the same figure are the 
weighted linear least-squares fit to the data taking into account the errors in both the
colours and the magnitudes. The results of the fits are given in Table \ref{table-4}.
We considered a source to have shown colour variations if the null hypothesis for correlation is less than 0.001. With this criterion, we found 20 sources to have shown colour variations, all of which were found to show a BWB trend.

\begin{figure*}
\centering
\includegraphics[width=4cm]{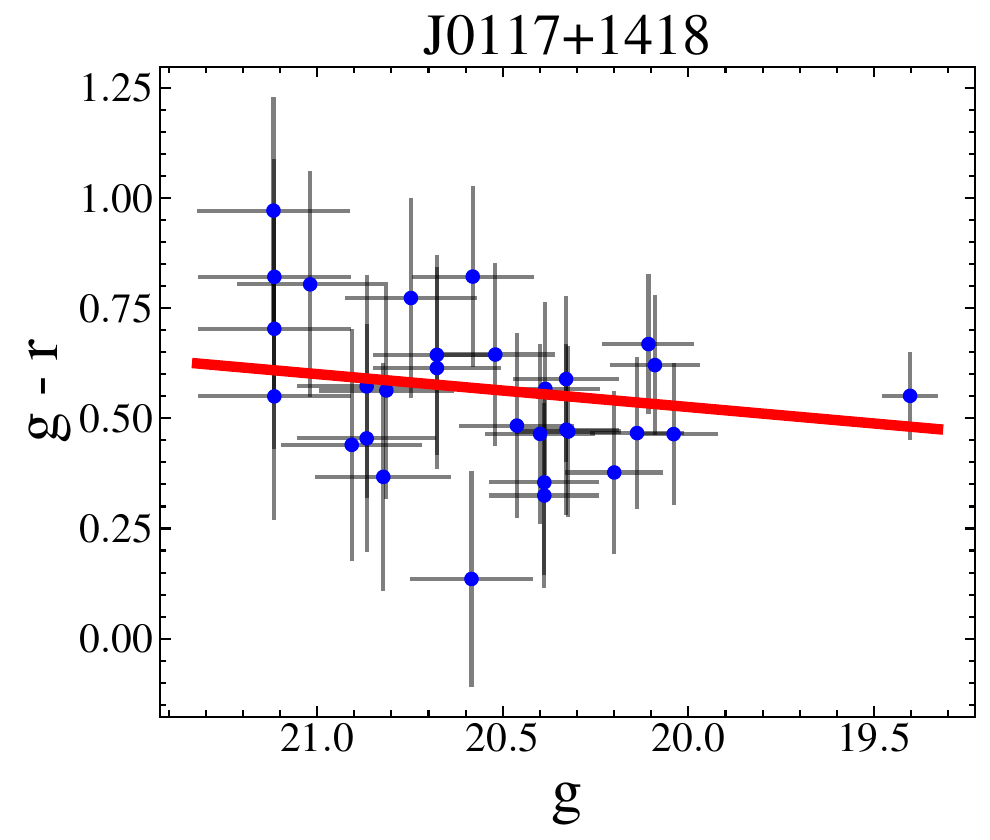}
	\includegraphics[width=4cm]{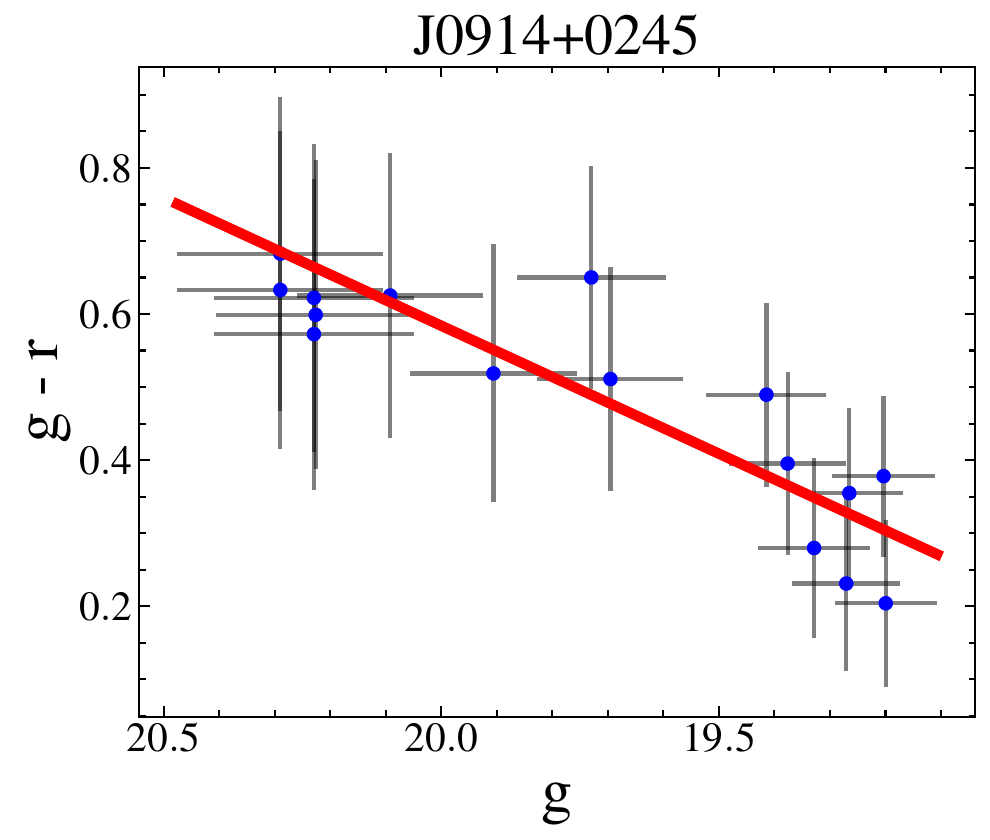}
	\includegraphics[width=4cm]{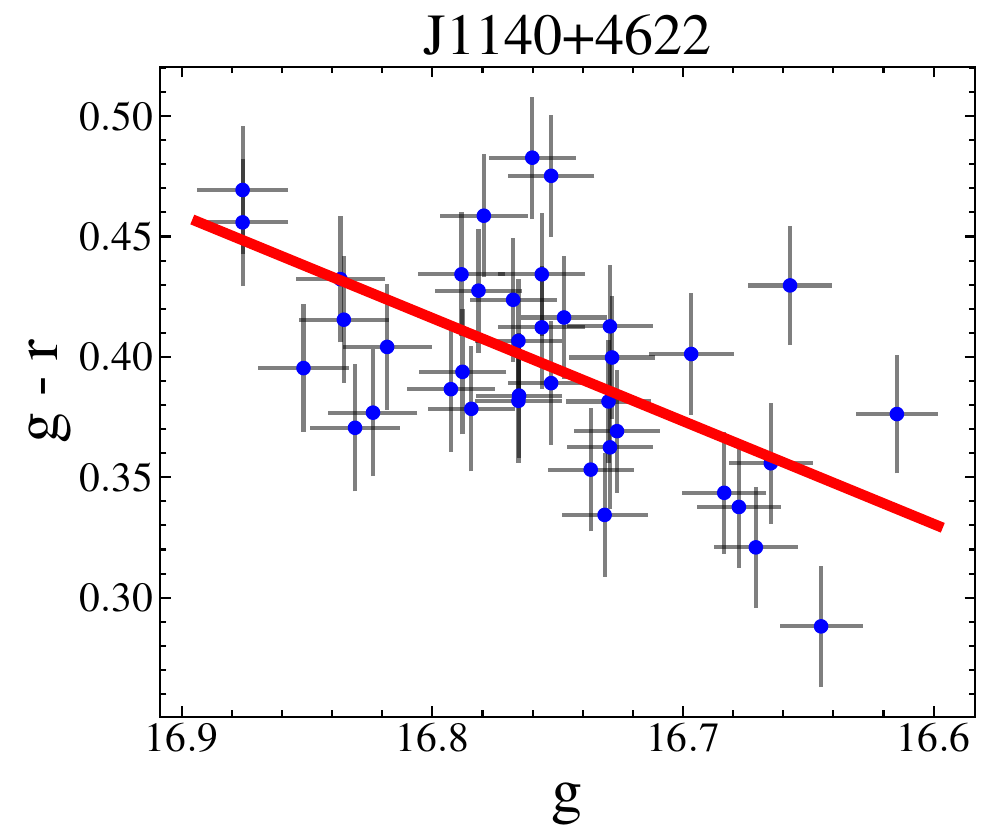}
	\includegraphics[width=4cm]{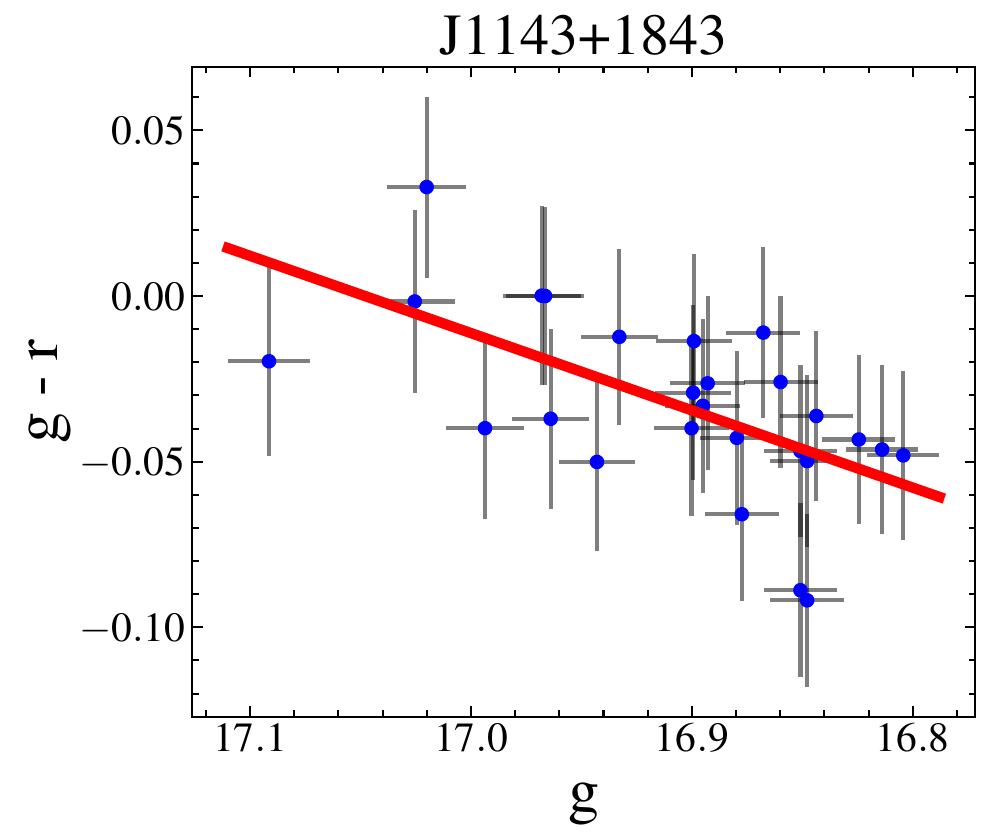}
	\includegraphics[width=4cm]{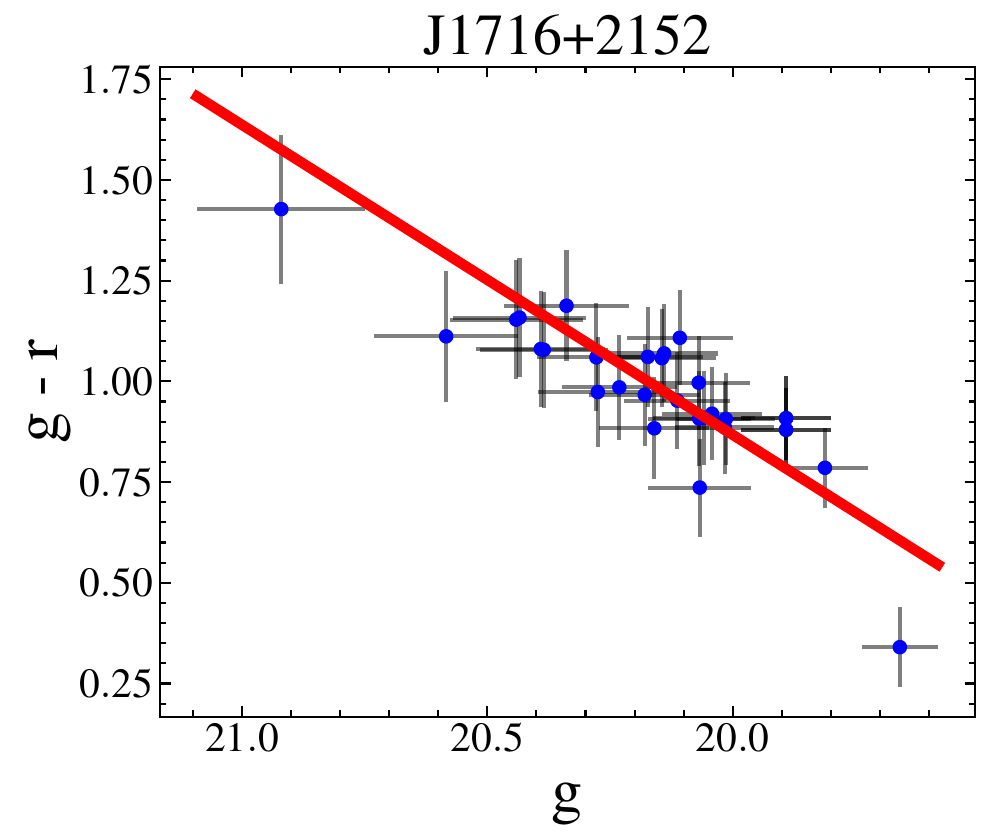}
\vbox{
    \includegraphics[width=4cm]{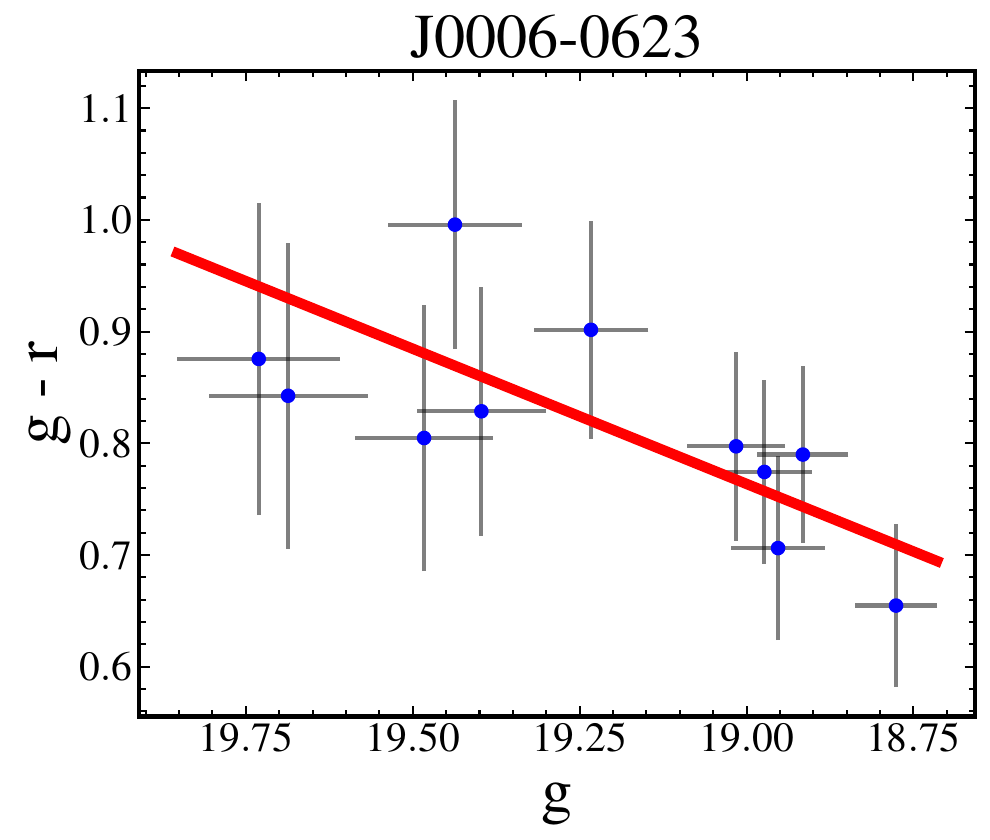}
	\includegraphics[width=4cm]{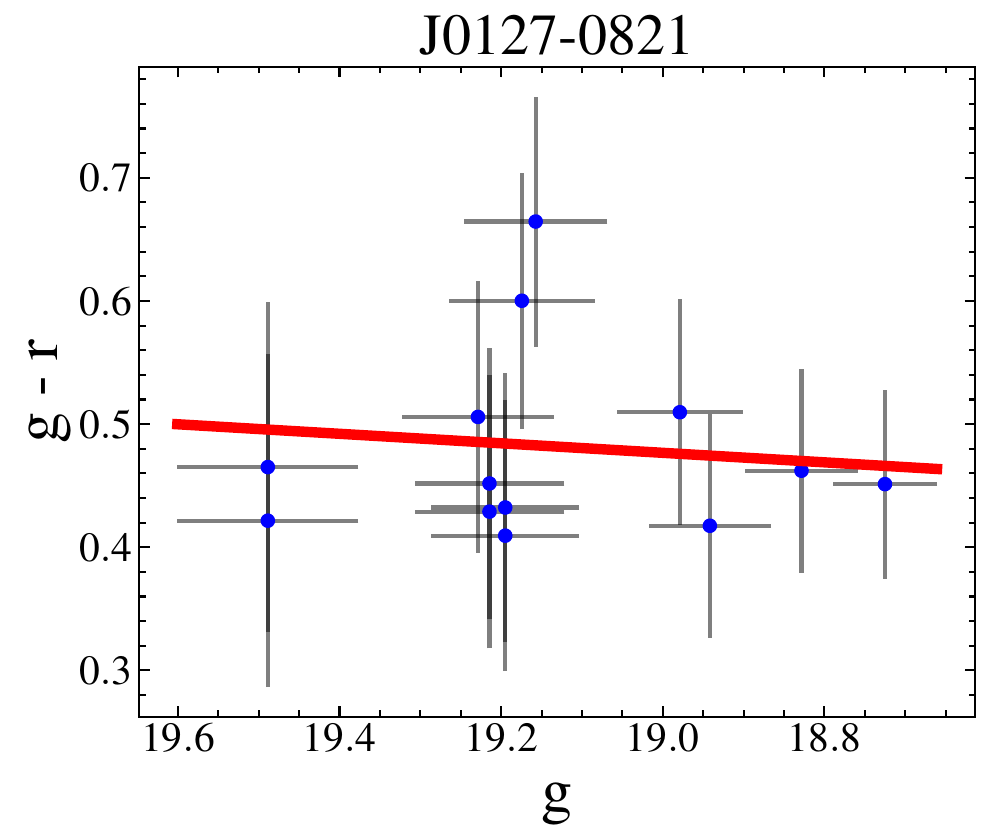}
	\includegraphics[width=4cm]{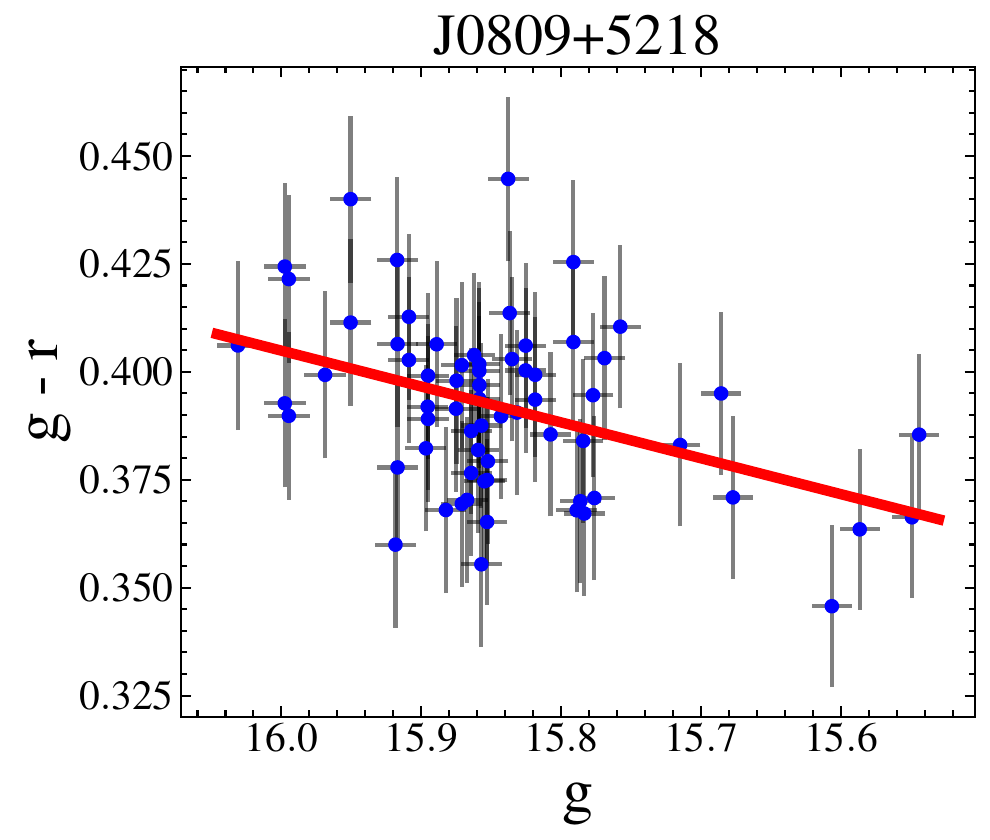}
	\includegraphics[width=4cm]{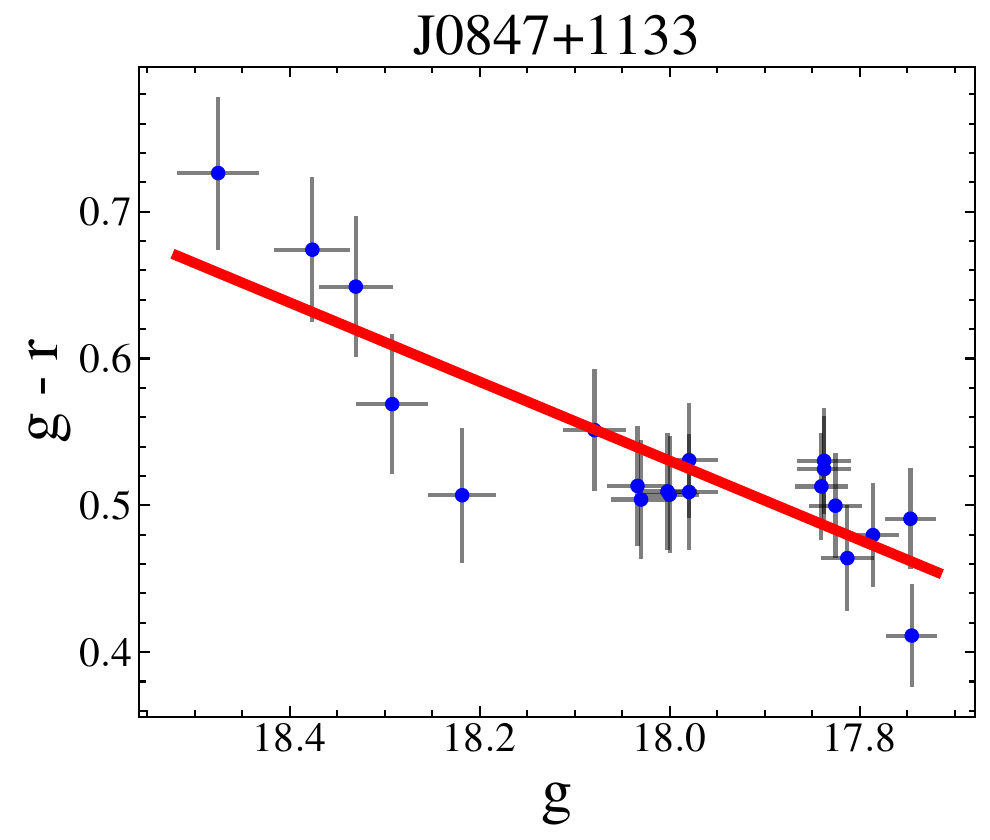}
	\includegraphics[width=4cm]{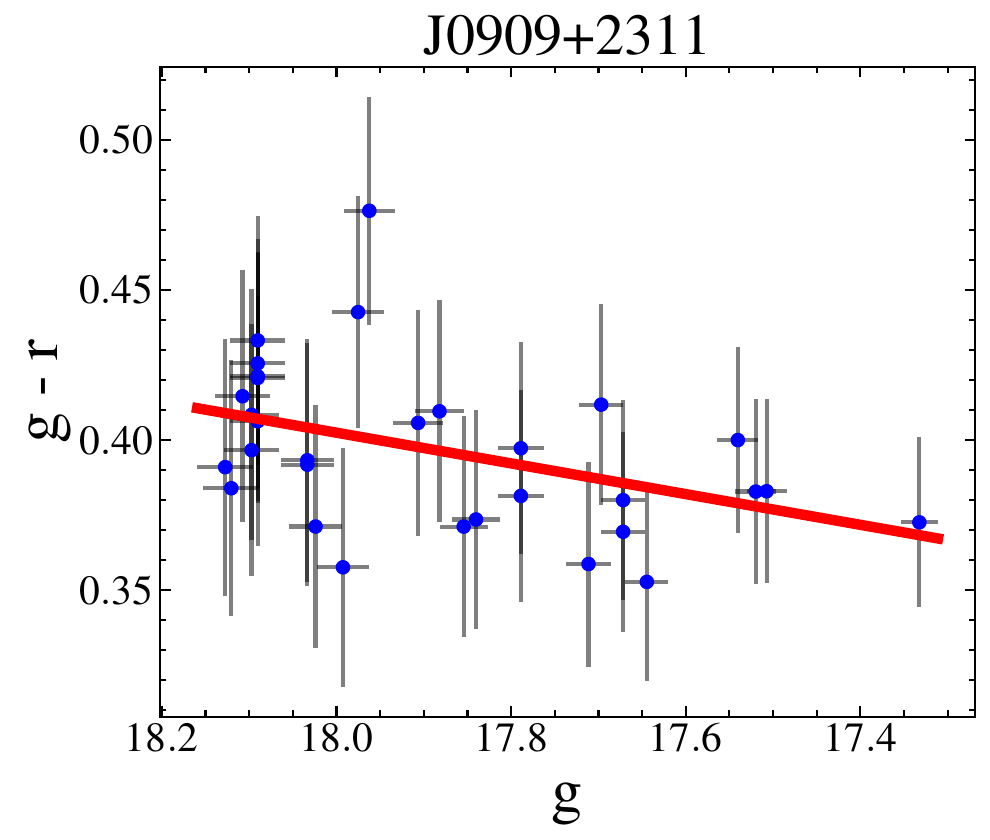}
	\includegraphics[width=4cm]{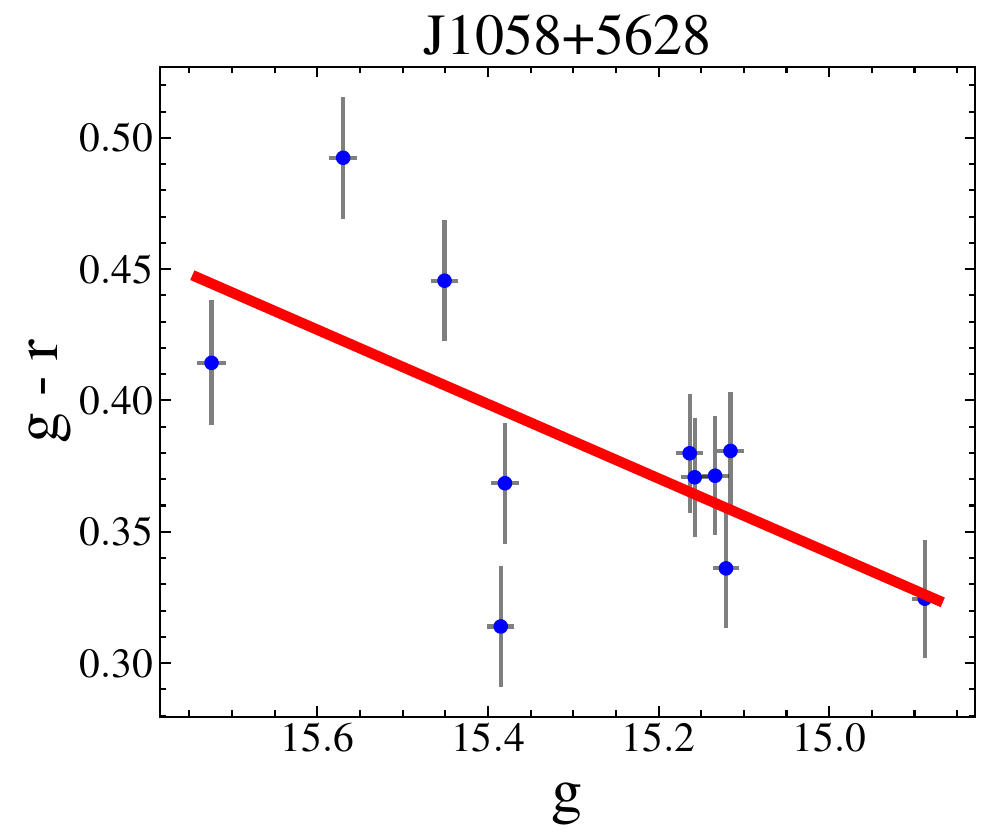}
	\includegraphics[width=4cm]{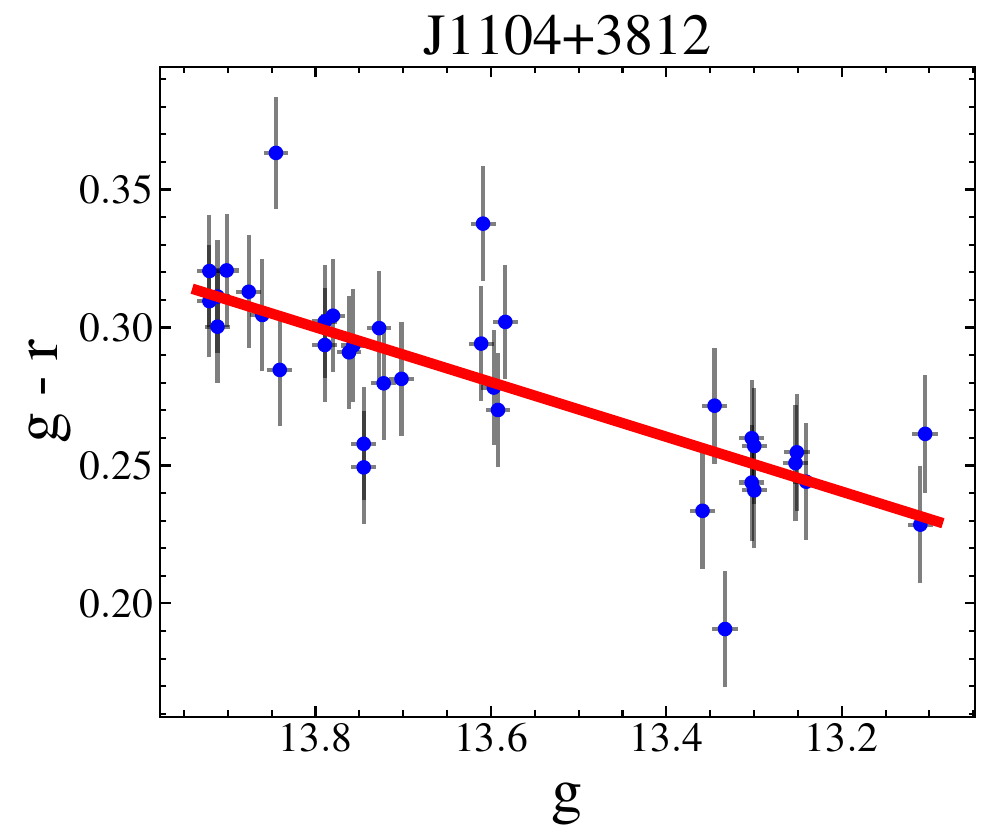}
	\includegraphics[width=4cm]{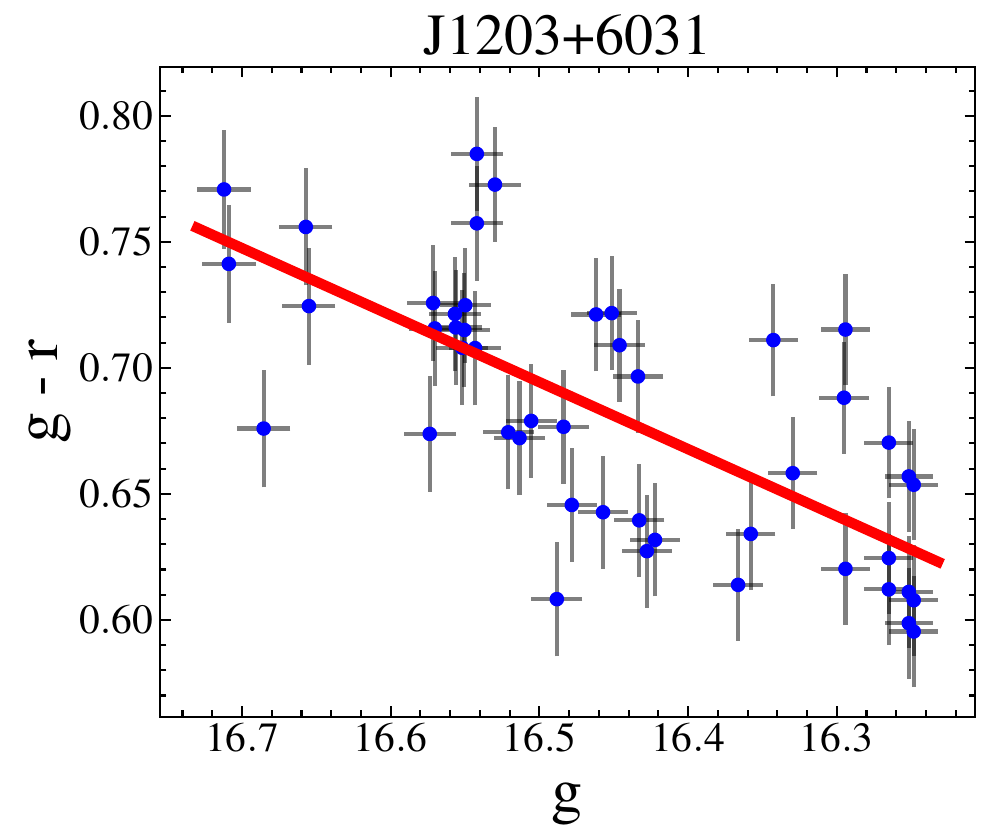}
	\includegraphics[width=4cm]{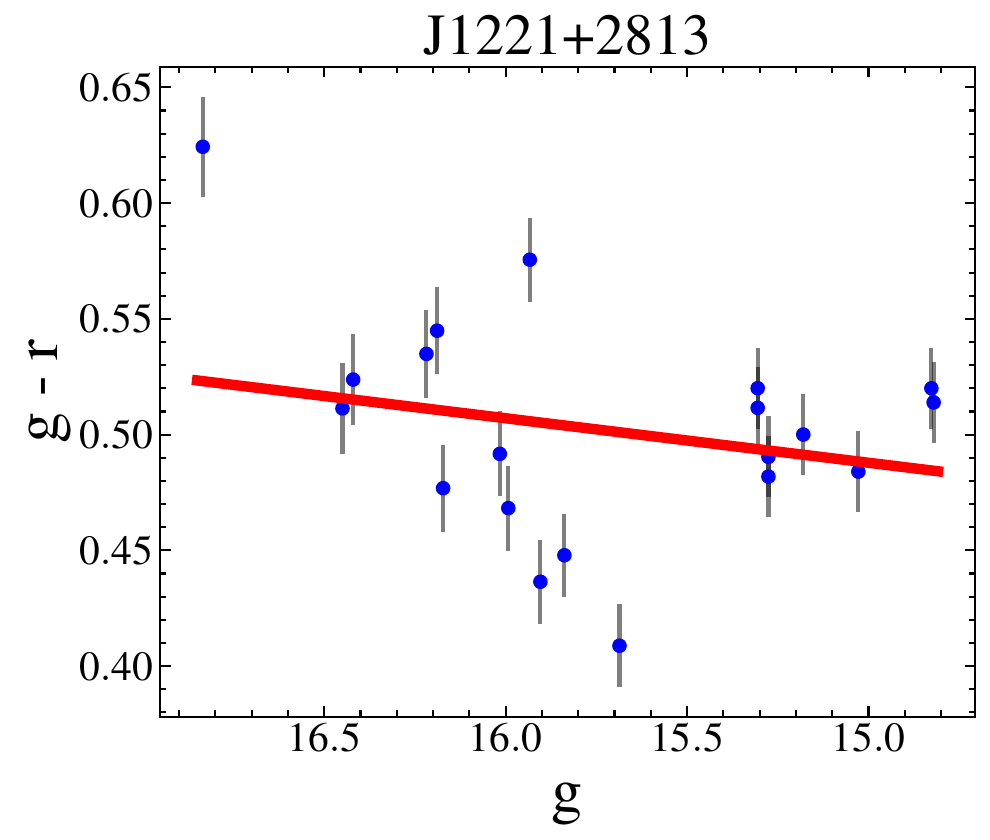}
	\includegraphics[width=4cm]{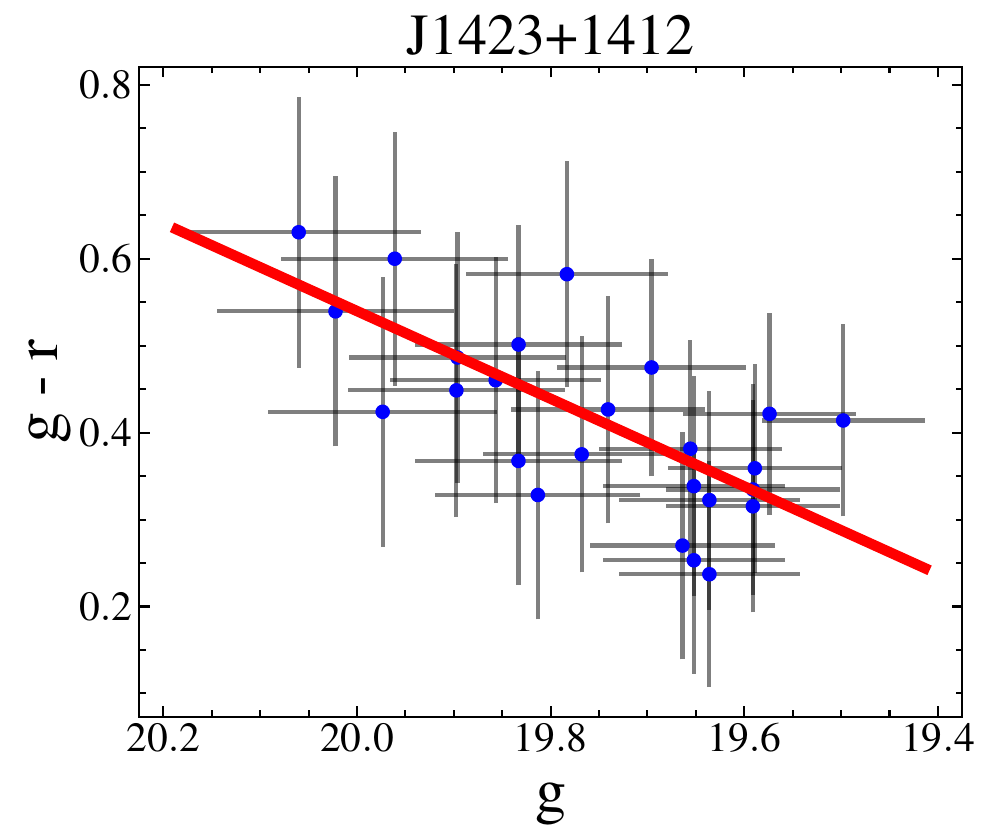}
	\includegraphics[width=4cm]{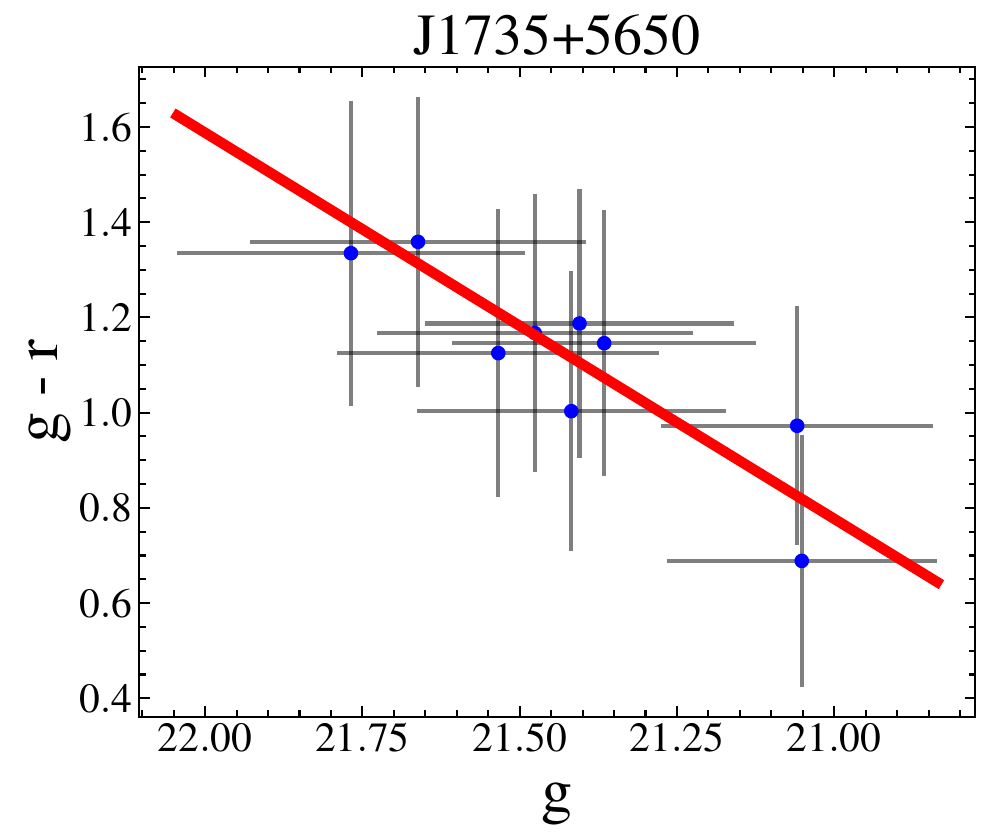}
	\includegraphics[width=4cm]{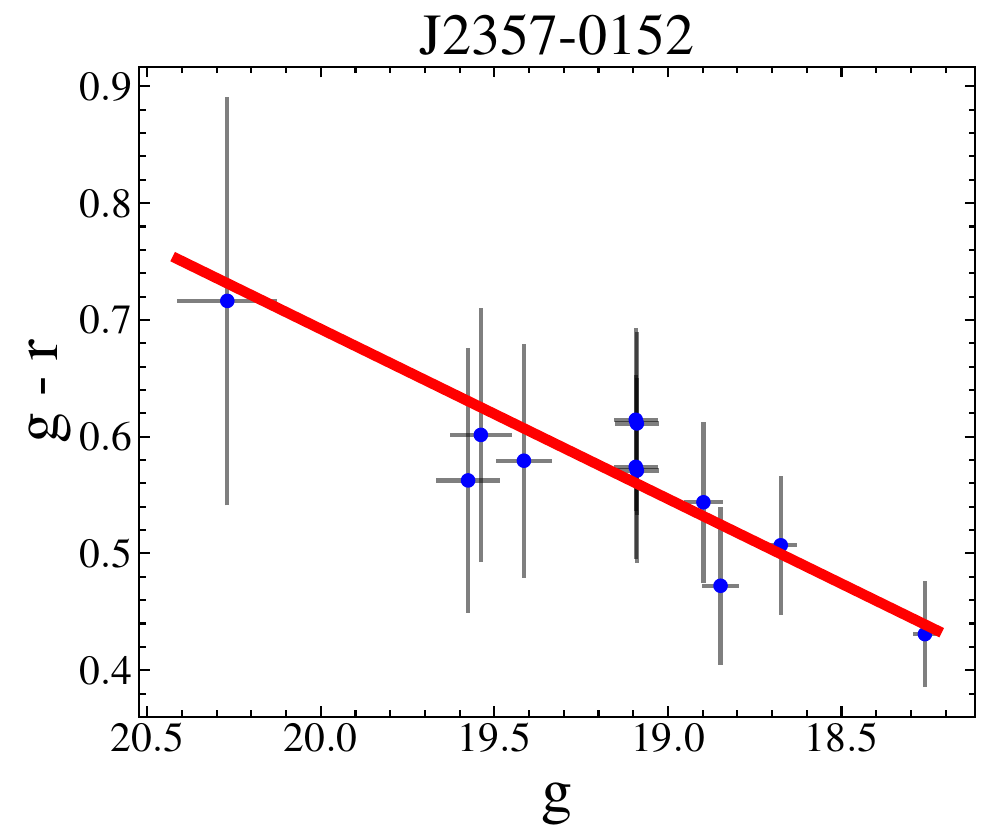}
}
    \caption{Colour-magnitude diagram for the sample of  FSRQs (top two rows) and BL Lacs (bottom four rows). Other information are same as in Figure~\ref{figure-3}.}
    \label{figure-6}
\end{figure*}

The variation of $g-r$ colour with $g$-band brightness was also investigated for blazars. The colour-magnitude diagrams for FSRQs and BL Lacs are shown in Figure \ref{figure-6}. Imposing the requirement for a source to have a minimum of 5 near simultaneous measurements in $g$ and $r$ bands, we found 5 FSRQs and 12 BL Lacs. The results of the linear least squares fits to the colour-magnitude diagrams are given in Table \ref{table-4}. Here too, similar to CSOs, we considered a source to have shown colour variations if the null hypothesis probability for no correlation is smaller than 0.001. With this criterion, of the 5 FSRQs, we found 3 sources to show a BWB trend. Among the 12 BL Lacs,  6 BL Lacs showed a distinct BWB trend. Considering the 9 CSOs for the matched blazar sample, 6 sources showed a BWB trend. The remaining sources did not show any significant correlation. 

\begin{table*}
\centering
\caption{Mean fractional variability amplitude (F$_{var}$) for the sample of CSOs.}
\label{table-3}
\begin{tabular}{cccccccc}
\hline
Type & No.&{$F_{var,g}$} & {$F_{var,r}$} & {$F_{var,i}$}\\
\hline
CSO (full)  & 21 & 0.084$\pm$0.001   & 0.070$\pm$0.001 &  0.066$\pm$0.001 \\
CSO (galaxies)  & 17 & 0.084$\pm$0.001 &  0.066$\pm$0.001 &  0.065$\pm$0.001 \\
CSO (quasars)  & 4 & 0.087$\pm$0.002 & 0.088$\pm$0.002 &  0.073$\pm$0.003 \\
\hline 
\end{tabular} 
\end{table*}

\begin{table*}
\centering
\caption{The mean $F_{var}$ for the matched sub-sample of CSOs, FSRQs and BL Lacs. The number of sources in different categories are provided in brackets.}
\label{table-5}
\small
\begin{tabular}{llll}
\hline 
Types &  $F_{var,g}$ & $F_{var,r}$ & $F_{var,i}$\\ 
\hline 
CSO (7)& 0.081$\pm$0.006 & 0.078$\pm$0.001 & 0.065$\pm$0.002 \\
FSRQ (5)     & 0.236$\pm$0.003 & 0.188$\pm$0.001 & 0.136$\pm$0.004 \\
BL Lac (10)      &0.251$\pm$0.001  & 0.228$\pm$0.001 & 0.185$\pm$0.001\\
\hline
\end{tabular} 
\end{table*}

\begin{table*}
\centering
	\caption{Results of correlation analysis between variations in the optical 
($g$ $-$ $r$) colour against the optical g-band brightness for 25 CSOs, 5 FSRQs and 12 BL Lacs. Here, R and p are the linear correlation coefficient 
and the probability of no correlation respectively.}
	\label{table-4}
\small
\begin{tabular}{llcc}
\hline 
Name  &  Slope  & R & p \\ 
\hline
\multicolumn{4}{c}{CSOs}\\
J0119+3210	&1.59$\pm$0.11  &0.79& $<$ 10$^{-5}$\\
J0131+5545  &1.26$\pm$0.20  &0.69& $<$ 10$^{-5}$\\ 
J0832+1832  &1.02$\pm$0.07  &0.92&	$<$ 10$^{-5}$ \\
J0906+4124  &0.90$\pm$0.13  &0.83&  $<$ 10$^{-5}$\\
J0909+1928  & 0.99$\pm$0.12 &0.84&  $<$ 10$^{-5}$\\ 
J1025+1022  &0.85$\pm$0.13  &0.87&  $<$ 10$^{-5}$\\  
J1111+1955  &0.86$\pm$0.17  &0.77&	0.0008\\
J1148+5924  & 0.84$\pm$0.10 &0.74& $<$ 10$^{-5}$\\
J1158+2450	&1.39$\pm$0.08  &0.89&	$<$ 10$^{-5}$\\
J1159+5820  & 0.27$\pm$0.14 &0.39& 0.1300\\
J1205+2031  & 0.56$\pm$0.29 &0.20& 0.4200\\
J1220+2916  & 1.16$\pm$0.14 &0.64& $<$ 10$^{-5}$\\
J1234+4753  &0.29$\pm$0.07  &0.46& 0.0012\\
J1244+4048  &0.05$\pm$0.09  &0.05& 0.7300\\
J1326+3154	& 1.15$\pm$0.11 &0.93& $<$ 10$^{-5}$\\
J1407+2827  & 0.10$\pm$0.11 &0.17& 0.4300\\
J1511+0518  & 1.51$\pm$0.27 &0.71& 0.0001 \\
J1559+5924  &1.12$\pm$0.12  &0.70& $<$ 10$^{-5}$\\
J1816+3457  & 1.05$\pm$0.05 &0.94& $<$ 10$^{-5}$\\
J1915+6548  & 1.19$\pm$0.07 &0.77& $<$ 10$^{-5}$\\
J1928+6815  & 0.91$\pm$0.07 &0.83& $<$ 10$^{-5}$\\
J1602+5243 &1.02$\pm$0.12   &0.64& $<$ 10$^{-5}$\\
J1945+7055 & 1.04$\pm$0.08  &0.92& $<$ 10$^{-5}$\\
J2022+6136 &1.12$\pm$0.06   &0.89& $<$ 10$^{-5}$\\
J2355+4950 & 1.01$\pm$0.06  &0.96& $<$ 10$^{-5}$\\ 
\hline 
\multicolumn{4}{c}{FSRQs}\\  
J0117+1418  &0.07$\pm$0.06&		0.35&	0.06\\
J0914+0245	&0.35$\pm$0.06&		0.88&	$<$ 10$^{-5}$\\
J1140+4622	&0.43$\pm$0.10&		0.53&	0.0005\\
J1143+1843  &0.23$\pm$0.06&		0.59&	0.0020\\
J1716+2152	&0.81$\pm$0.02&		0.85&   $<$ 10$^{-5}$ \\
\hline 
\multicolumn{4}{c}{BL Lacs}\\
J0006$-$0623	&0.24$\pm$0.07  & 0.68&	0.0200\\
J0127$-$0821    &0.04$\pm$0.09	& 0.02&	0.9600\\
J0809+5218	&0.08$\pm$0.02&	0.41&	0.0007\\
J0847+1133	&0.26$\pm$0.04&	0.86&	$<$ 10$^{-5}$\\
J0909+2311 	&0.05$\pm$0.02&	0.43&	0.0200\\
J1058+5628	&0.14$\pm$0.06& 0.64&	0.0300\\
J1104+3812	&0.09$\pm$0.01&	0.75&	$<$ 10$^{-5}$\\
J1203+6031 	&0.26$\pm$0.04&	0.69&	$<$ 10$^{-5}$\\
J1221+2813  &0.02$\pm$0.02& 0.31&	0.1800\\
J1423+1412	&0.50$\pm$0.11&	0.68&	0.0001\\
J1735+5650	&0.81$\pm$0.16&	0.88&	0.0020\\
J2357$-$0152&0.15$\pm$0.02&	0.88&	0.0002\\
\hline
\end{tabular}
\end{table*}

\section{Discussion}
Flux variability has been recognised as a defining property of AGN. They have been studied extensively across wavelengths, either as individual sources or as a sample to understand the cause of flux variations in them. However, studies of optical flux variability of compact steep-spectrum sources, including CSOs, have been limited. In this work, we used the recent compilation of a bona fide
sample of CSOs by \cite{Kiehlmann2023b} to examine their optical variability characteristics. The CSOs are recognised as a distinct
class of jetted AGN. Their reasonably symmetric morphology, unlike the core-dominated sources with a core-jet morphology, points to them being viewed at large angles to the line of sight. These sources are thus in sharp contrast to blazars, which have dominant cores and are viewed with their jet axes inclined at small angles to the line of sight. 
The primary objective of this work was to characterise the optical variability nature of CSOs and look for similarities and/or differences in their 
variability properties relative to blazars. It is important to characterise the optical 
variability properties of CSOs, because, in contrast to blazars, which have been well studied over the past decades  \citep{2009MNRAS.399.1357S,2012ApJ...756...13B,2017ApJ...835..275R,Paliya2017}, the CSOs lack a comparable level of investigation. To address this, we used the light curves with large temporal coverage in multiple optical filters offered by the ZTF database.

From the temporal analysis of the sample of CSOs as a whole, we found that $\sim$ 76\%, $\sim$ 87\% and $\sim$ 78\% CSOs are variable in $g$, $r$ and $i$ bands respectively. Furthermore, $\sim$55\% show variability across all three bands. We found the mean $F_{var}$ to be 0.084$\pm$0.001, 0.070$\pm$0.001 and 0.066$\pm$0.001 in $g$, $r$ and $i$ bands, respectively. This is similar to the variations seen in the subset of CSOs that were selected to match with the blazars. The detection of significant flux variations indicates the jet origin of the optical radiation detected from these objects. Moreover, the observation of a BWB trend, typically seen in jet-dominated sources, from most of the CSOs supports this hypothesis. 

Both FSRQs and BL Lacs are variable and tend to show greater fractional variability amplitude compared to CSOs with estimated $F_{var}$ values found to be larger than those of CSOs.
This is because any perturbations (leading to flux variations) in blazars jets get amplified due to strong Doppler boosting effects. On the other hand, CSOs are thought to be viewed at large viewing angles, hence modest Doppler boosting, and thus relatively low level of variability.


In addition to flux variations, blazars have also exhibited strong colour variations. In FSRQs, which are mostly low-frequency peaked blazars \citep[e.g.,][]{2010ApJ...716...30A, Fan_2016}, 
the observed optical emission is likely to be a combination of the redder non-thermal jet radiation and the bluer thermal emission from the accretion disk \citep{2008bves.confE...9P}. Therefore, the observed colour variation and its relation to the optical brightness could be governed by the complex interplay between the accretion disk and the relativistic jet. The optical emission from BL Lacs, on the other hand, is usually dominated by the synchrotron radiation from the jet, so the observed colour variations would be driven by the jet activity \citep[e.g.,][]{Marscher1985, Marscher_2014}. Also, larger variations at shorter wavelengths can happen due to the injection of fresh electrons having an energy distribution harder than the earlier cooler electrons \citep{Krik1998, mastichiadis_kirk_2002},  leading to an increase in flux with a BWB trend. The changes in the Doppler factor can also lead to the observation of a BWB behavior \citep{Villata2004}. Such a change in the Doppler factor is expected due to geometric effects leading to changes in the viewing angle of the relativistically moving plasma blob to the observer \citep{Papadakis2007}.

\section{Conclusions}
We used the ZTF database to characterise the long-term optical flux
and colour variability of a bona fide sample of 38 CSOs. To check how the optical variability properties of CSOs compare with the beamed population of blazars, we also carried out a comparative analysis of the flux and colour variability characteristics of a matched and limited sub-sample of 9 CSOs, 5 FSRQs, and 12 BL Lacs. This is the first
systematic study aimed at characterising the long-term optical variability
characteristics of a bona fide sample of CSOs. The results of the work are 
summarized below:

\begin{enumerate}
\item There are 21 CSOs that exhibit flux variations in all $g$, $r$ and $i$ bands, while 6 sources are variable in the $g$ and $r$ bands with their $i$-band data unavailable. Additionally, 9 sources show variability in at least one band, one source is a probable variable, and one is a non-variable. CSOs showed weak, low-amplitude long-term optical flux variations in all bands. Considering mean values, the fractional variability amplitude of variation is more in the shorter wavelength $g$ band compared to the $r$ and $i$ bands.
\item The CSOs associated with quasars show flux variations in the $g$, $r$, and $i$ bands with a fractional variability amplitude which tends to be higher compared to those associated with galaxies. This difference is seen to be most significant in the $r$ band. This is expected as CSO quasars are believed to be inclined at smaller angles to the line of sight than CSO galaxies. However, as there are only four quasars in this sample, sensitive observations of a large sample of CSO galaxies and quasars are required to explore this further.
\item All FSRQs and BL Lacs studied showed long-term optical variability. Here, too,
the mean fractional amplitude of flux variations is larger in $g$ band and lower in $r$ and $i$ bands with the trend $F_{var,g}$ $>$ $F_{var,r}$ $>$ $F_{var,i}$.  

\item In a matched sub-sample of a limited number of CSOs and blazars, 
the variations in CSOs are lower than FSRQs and BL Lacs. 
 Such reduced variability in
CSOs relative to blazars is due to the fact that CSOs are inclined at larger angles to the line of sight and the effects of relativistic beaming are small.

\item All the categories of sources, namely CSOs, FSRQs and BL Lacs showed a bluer when brighter chromatic behaviour. The similarity in the colour-magnitude behaviour between CSOs and blazars and the decreased amplitude of flux variations in CSOs relative to blazars point to relativistic jet origin of flux variations. The larger $F_{var}$ noticed for blazars is due to stronger relativistic beaming effects. 
\end{enumerate} 

\section{Acknowledgements}
We thank the journal referee for constructive suggestions that helped us improve the manuscript.  
This work is based on observations obtained with the Samuel
Oschin Telescope 48-inch and the 60-inch Telescope at the Palomar Observatory as part of the Zwicky Transient Facility project.
ZTF is supported by the National Science Foundation under Grant
No. AST-2034437 and a collaboration including Caltech, IPAC, the
Weizmann Institute for Science, the Oskar Klein Center at Stockholm University, the University of Maryland, Deutsches Elektronen Synchrotron and Humboldt University, the TANGO Consortium of
Taiwan, the University of Wisconsin at Milwaukee, Trinity College Dublin, Lawrence Livermore National Laboratories, and IN2P3,
France. Operations are conducted by COO, IPAC, and UW.
This research has made use of the NASA/IPAC Extragalactic
Database (NED) which is operated by the Jet Propulsion Laboratory,
California Institute of Technology, under contract with the National
Aeronautics and Space Administration.

\bibstyle{sn-basic}
\bibliography{reference.bib}

\end{document}